\documentclass[journal]{IEEEtran}
\usepackage{epsfig,graphics,epsf,color,amsmath,balance,cite,amsfonts,multirow,stfloats,algorithm,algorithmic,mathrsfs,amssymb,bm}
\usepackage[dvipdfm,unicode,colorlinks, linkcolor=black,anchorcolor=black,citecolor=black]{hyperref}
\newtheorem{theorem}{Theorem}
\newtheorem{lemma}{Lemma}
\newtheorem{proposition}{Proposition}
\newtheorem{remark}{Remark}

\begin{document}

\title{Joint Training and Reflection Pattern Optimization for Non-Ideal RIS-Aided Multiuser Systems}

\author{Zhenyao He,~\IEEEmembership{Graduate Student Member,~IEEE,} Jindan Xu,~\IEEEmembership{Member,~IEEE,}
Hong Shen,~\IEEEmembership{Member,~IEEE,}\\Wei Xu,~\IEEEmembership{Senior Member,~IEEE,} Chau Yuen, \IEEEmembership{Fellow,~IEEE,}~and~Marco Di Renzo,~\IEEEmembership{Fellow,~IEEE}

\thanks{
Zhenyao He, Hong Shen, and Wei Xu are with the National Mobile Communications Research Laboratory, Southeast University, Nanjing 210096, China (e-mail: \{hezhenyao, shhseu, wxu\}@seu.edu.cn).

Jindan Xu and Chau Yuen are with the School of Electrical and Electronics Engineering, Nanyang Technological University, Singapore 639798, Singapore (e-mail: jindan\_xu@sutd.edu.sg, chau.yuen@ntu.edu.sg).

Marco Di Renzo is with Universit\'e Paris-Saclay, CNRS, CentraleSup\'elec, Laboratoire des Signaux et Syst\`emes, 3 Rue Joliot-Curie, 91192 Gif-sur-Yvette, France (e-mail: marco.di-renzo@universite-paris-saclay.fr).
}}

\maketitle
\begin{abstract}
Reconfigurable intelligent surface (RIS) is a promising technique to improve the performance of future wireless communication systems at low energy consumption. To reap the potential benefits of RIS-aided beamforming, it is vital to enhance the accuracy of channel estimation. In this paper, we consider an RIS-aided multiuser system with non-ideal reflecting elements, each of which has a phase-dependent reflecting amplitude, and we aim to minimize the mean-squared error (MSE) of the channel estimation by jointly optimizing the training signals at the user equipments (UEs) and the reflection pattern at the RIS. As examples the least squares (LS) and linear minimum MSE (LMMSE) estimators are considered.
The considered problems do not admit simple solution mainly due to the complicated constraints pertaining to the non-ideal RIS reflecting elements.
As far as the LS criterion is concerned, we tackle this difficulty by first proving the optimality of orthogonal training symbols and then propose a majorization-minimization (MM)-based iterative method to design the reflection pattern, where a semi-closed form solution is obtained in each iteration.
As for the LMMSE criterion, we address the joint training and reflection pattern optimization problem with an MM-based alternating algorithm, where a closed-form solution to the training symbols and a semi-closed form solution to the RIS reflecting coefficients are derived, respectively. Furthermore, an acceleration scheme is proposed to improve the convergence rate of the proposed MM algorithms.
Finally, simulation results demonstrate the performance advantages of our proposed joint training and reflection pattern designs.
\end{abstract}

\begin{IEEEkeywords}
Reconfigurable intelligent surface (RIS), channel estimation, least squares (LS), linear minimum mean-squared error (LMMSE), reflection pattern, majorization-minimization (MM).
\end{IEEEkeywords}

\section{Introduction}
The deployment of reconfigurable intelligent surfaces (RISs) has drawn a lot of attention as a cost-effective promising solution for wireless communication networks \cite{RISoverview1,RISoverview2,RISoverview3,RISoverview4,RISoverview5,RISoverview6,W.Xu,J.Xuoverview}. An RIS usually consists of a large number of low-cost passive adjustable reflecting elements, each of which can be independently controlled to adjust the amplitude and/or the phase of the reflected signals such that the wireless transmission channels are sculpted to fulfill various design goals.

Recently, considerable innovative contributions have been devoted to optimize RIS-aided wireless communications. The joint optimization of the transmit beamforming at the transmitter and the reflection beamforming at the RIS has been studied to maximize the received signal-to-noise ratio (SNR) \cite{SINRmax} and to minimize the total transmit power \cite{Pmin} for a single-user RIS-assisted multiple-input single-output (MISO) system. In \cite{ratemax} and \cite{EEmax}, the authors considered maximizing the sum rate and the energy efficiency for a downlink RIS-aided multiuser MISO system, respectively.
The secrecy rate maximization problem for RIS-assisted multi-antenna communications has been studied in \cite{HShenSecrecyrate1,Secrecyrate2} from the physical-layer security perspective.
In \cite{RISFD}, the authors studied the sum rate maximization problem for RIS-aided full-duplex communications \cite{FD1,FD2}.
Moreover, practical low-resolution RIS phase shifts were further considered in single-user \cite{discrete1} and multiuser \cite{discrete2} systems.
In \cite{hardwareimpairments}, the received SNR was maximized for an RIS-aided single-user system by considering the impact of practical transceiver hardware impairments.
Different from the above works focusing on flat-fading channels, an RIS-aided orthogonal frequency division multiplexing (OFDM) system over frequency-selective channels was studied in \cite{OFDM1,OFDM2,OFDM3,onoffgroup}.
As revealed in these works, an RIS is proved beneficial for improving the performance of wireless communications.

To fully achieve the benefits of RIS-aided communications, it is necessary to obtain accurate channel state information (CSI), which, however, turns out to be technically challenging \cite{separateCE}, since an RIS cannot transmit or receive pilots to assist the channel estimation. To overcome this difficulty, a cascaded channel estimation method was investigated in \cite{onoff,fullpattern,CE1,JointXTheta,CE2,CEsparsity3,KFtransform}, which only requires pilot signals at the transmitter.
Concretely, an ``on-off'' reflection pattern design, which turns on one RIS element at a time, was first developed in \cite{onoff} for the cascaded channel estimation.
Subsequently, it is found that the channel estimation performance of RIS-aided systems can be enhanced by turning on all the RIS elements during the training phase and configuring the reflection pattern appropriately.
In this regard, the authors of \cite{fullpattern,CE1} optimized the RIS reflection pattern to minimize the mean-squared error (MSE) of the least squares (LS) channel estimator.
By exploiting the channel statistics knowledge in RIS-aided systems, the authors of \cite{CE2} demonstrated that the linear minimum MSE (LMMSE) channel estimator can achieve a better MSE performance. Then, in \cite{JointXTheta}, the joint optimization of the training sequence at the transmitter and the reflection pattern at the RIS was analyzed for the LMMSE estimator.
In addition, the channel sparsity was exploited in \cite{CEsparsity3} to assist the cascaded channel estimation. In \cite{KFtransform}, a more complex RIS-aided multi-cell system was considered and the corresponding CSI acquisition method was investigated. On the other hand, the dimension of the cascaded channel in RIS-aided systems could be quite high due to the large number of RIS reflecting elements, thus leading to excessive training overhead. Several efforts have been made to address this issue \cite{onoffgroup,KFtransform,lowoverhead,twotimescale1,twotimescale2}. For example, the authors of \cite{onoffgroup} proposed a group-wise channel estimation method, by partitioning all the RIS elements into several groups, where each group consists of a set of neighboring RIS elements sharing a common reflection coefficient, thus reducing the effective number of RIS elements and the training overhead. A novel two-timescale CSI-based protocol was recently proposed \cite{twotimescale1,twotimescale2}, where the beamforming at the base station (BS) was designed based on the instantaneous CSI while the reflection coefficients at the RIS were optimized based on the slow time-varying long-term CSI. As a result, given a fixed RIS reflection pattern that remains unchanged within several coherence blocks, the dimension of the instantaneous channel that needs to be estimated in each block is independent of the number of RIS elements, so that the training overhead can be significantly reduced.

Most of the existing works on the channel estimation for RIS-assisted systems assume unit amplitude signal reflection, i.e., the magnitude of the reflection coefficient of each RIS element is always one, regardless of the phase shift, so as to maximize the reflected signal power. However, it was revealed in \cite{realisticRIS1,realisticRIS2,realisticRIS3,realisticRIS4} that such assumption is actually ideal and the magnitude of the reflection coefficient of a practical RIS element is dependent on the reflection phase \cite{RISPracticalModel}. In other words, one cannot adjust the reflection phase while maintaining the unit reflection magnitude at the same time.

Motivated by the above facts, we investigate the joint training and reflection pattern design for channel estimation in an RIS-aided multiuser system considering the LS and LMMSE channel estimators, by explicitly taking into account the characteristics of non-ideal RIS elements.
In such cases, most methods developed in prior related works, e.g., in \cite{fullpattern,CE1,JointXTheta}, that rely on an ideal RIS with unit-modulus reflection coefficient cannot be applied to the considered problems. In fact, the considered phase-dependent amplitude configuration makes the corresponding optimization nontrivial and challenging. To tackle these challenges, our main contributions are summarized as follows:
\begin{itemize}
  \item As for the LS channel estimator, we reveal that the optimal training sequences at the user equipments (UEs) are independent of the reflection pattern at the RIS and are orthogonal to each other.
      Then, we develop a majorization-minimization (MM)-based iterative algorithm to address the reflection pattern design, where a semi-closed form solution is obtained in each iteration.
  \item As for the LMMSE channel estimator, the corresponding optimization problem turns out to be more difficult to solve. To obtain a tractable and also high-quality solution, we optimize the pilots and the reflection pattern in an alternating way. In particular, by invoking the MM technique, we obtain a closed-form solution and a semi-closed form solution for the pilot sequence and the reflection pattern, respectively.
  \item The convergence of the two proposed joint training and reflection pattern optimization designs is theoretically proved. Furthermore, we propose an accelerated strategy for the developed algorithms in order to significantly reduce the number of iterations required for reaching the convergence and reduce the overall computational complexity.
\end{itemize}

The rest of this paper is organized as follows. In Section~\uppercase\expandafter{\romannumeral2}, the RIS-aided multiuser communication system and the corresponding LS and LMMSE channel estimation methods are introduced. Section~\uppercase\expandafter{\romannumeral3} and Section~\uppercase\expandafter{\romannumeral4} present the proposed MM-based joint training and reflection pattern designs for the LS and LMMSE channel estimators, respectively. In Section~\uppercase\expandafter{\romannumeral5}, we propose an acceleration scheme to improve the convergence rate of the proposed MM algorithms. Simulation results are shown in Section~\uppercase\expandafter{\romannumeral6}. Finally, conclusions are provided in Section~\uppercase\expandafter{\romannumeral7}.

\textit{Notations}: Vectors and matrices are denoted by boldface lower-case and boldface upper-case letters, respectively. $\mathbb R$ and $\mathbb C$ denote the sets of real and complex numbers, respectively. The superscripts $(\cdot)^T$, $(\cdot)^*$, and $(\cdot)^H$ denote the transpose, the conjugate, and the conjugate transpose operations, respectively. $\|\cdot\|_F$ and $\text{Tr}[\cdot]$ denote the Frobenius norm and the trace of a matrix, respectively. $\|\cdot\|$ denotes the $\ell_2$ norm of a vector. $|\cdot |$, $\arg(\cdot)$, and $\mathcal R\{\cdot\}$ return the modulus, the phase, and the real part of the input complex number, respectively. $\otimes$ and $\odot$ stand for the Kronecker product and the Hadamard product, respectively. $[\cdot]_{i:j,p:q}$ returns the submatrix formed by the elements from the $i$-th to the $j$-th rows and from the $p$-th to the $q$-th columns of the input matrix. $\text{vec}(\cdot)$ is the vectorization operation. $\mathbb{E}\{\cdot\}$ represents the expectation operation. $\mathbf{a} \thicksim \mathcal{CN}(\mathbf{\bar a},\mathbf \Sigma)$ means that the vector $\mathbf{a}$ follows a circularly symmetric complex Gaussian distribution with mean $\mathbf{\bar a}$ and covariance matrix $\mathbf \Sigma$. $\text{diag}\{\cdot\}$ returns a diagonal matrix with diagonal elements being the entries of the input vector. $\mathbf I_N$ denotes the identity matrix of size $N \times N$. $\lambda_\text{max} (\mathbf A)$ returns the largest eigenvalue of matrix $\mathbf A$. $\mathcal O(\cdot)$ denotes the big-O computational
complexity notation.

\section{System Model and Problem Formulation}
In this section, we first describe the model and the channel estimation criteria of the considered RIS-assisted multiuser MISO system, and then introduce the problem formulation for the joint design of the uplink training symbols and the RIS reflection coefficients.

\subsection{System Model}
\begin{figure}[t]
\centering
\includegraphics[width=0.45\textwidth]{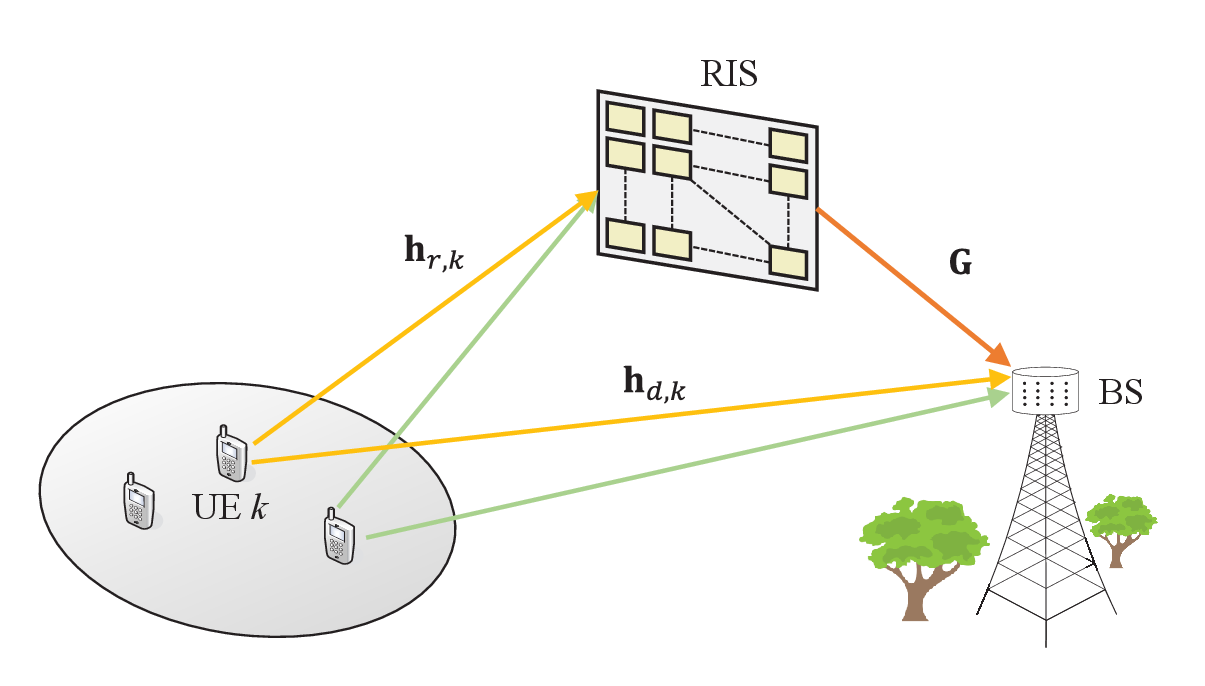}
\caption{The considered RIS-aided uplink multiuser communication system.}
\label{fig1:systemmodel}
\end{figure}

We consider the RIS-assisted uplink multiuser wireless communication system in Fig. \ref{fig1:systemmodel}, which consists of a BS with $L$ receive antennas, an RIS with $M$ reflecting elements, and $K$ single-antenna UEs.
Let $\mathbf{G} \in \mathbb{C}^{L \times M}$, $\mathbf{h}_{r,k} \in \mathbb{C}^{M \times 1}$, $\mathbf{h}_{d,k} \in \mathbb{C}^{L \times 1}$ denote the channel between the RIS and the BS, the channel between the $k$-th UE and the RIS, and the channel between the $k$-th UE and the BS, respectively, which are assumed to follow the Rayleigh fading model.
Denote the stacked channels of the UEs-RIS link and the UEs-BS link by $\mathbf H_r \triangleq [\mathbf h_{r,1}, \cdots, \mathbf h_{r,K}] \in \mathbb{C}^{M \times K}$ and $\mathbf H_d \triangleq [\mathbf h_{d,1}, \cdots, \mathbf h_{d,K}] \in \mathbb{C}^{L \times K}$, respectively.
Moreover, the reflection coefficient matrix at the non-ideal RIS is denoted by $\mathbf{\Phi} = \text{diag} \left\{[\phi_1, \cdots, \phi_M]^T \right\}$, with $\phi_m$ denoting the reflection coefficient of the $m$-th element at the RIS.

\begin{figure}[t]
\begin{center}
\epsfxsize=7.0in\includegraphics[scale=0.5]{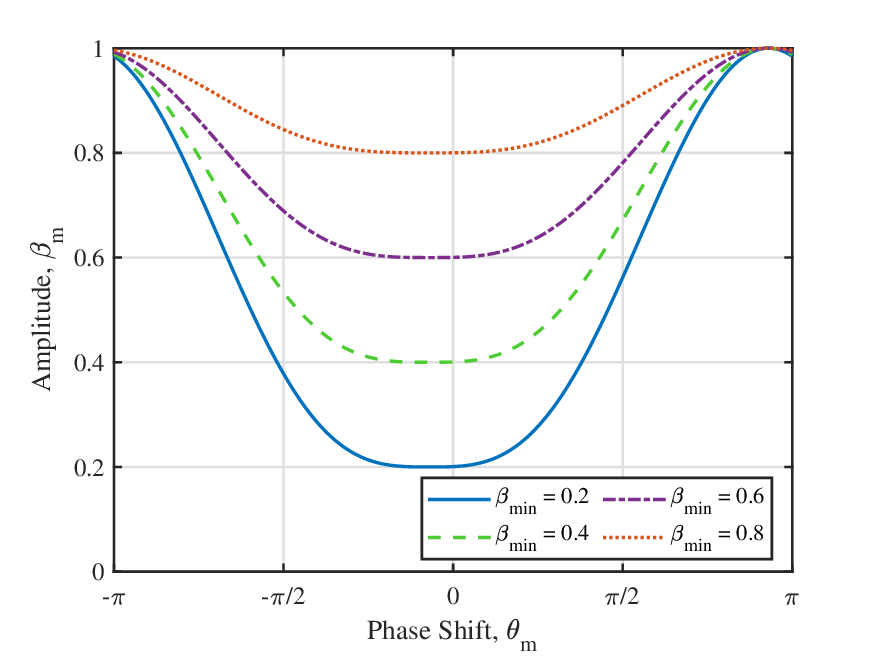}
\caption{The considered practical phase shift model \cite{RISPracticalModel}.}
\label{fig:fittedcurve}
\end{center}
\end{figure}

\begin{figure*}[t]
\centering
\epsfxsize=7.0in\includegraphics[scale=0.6]{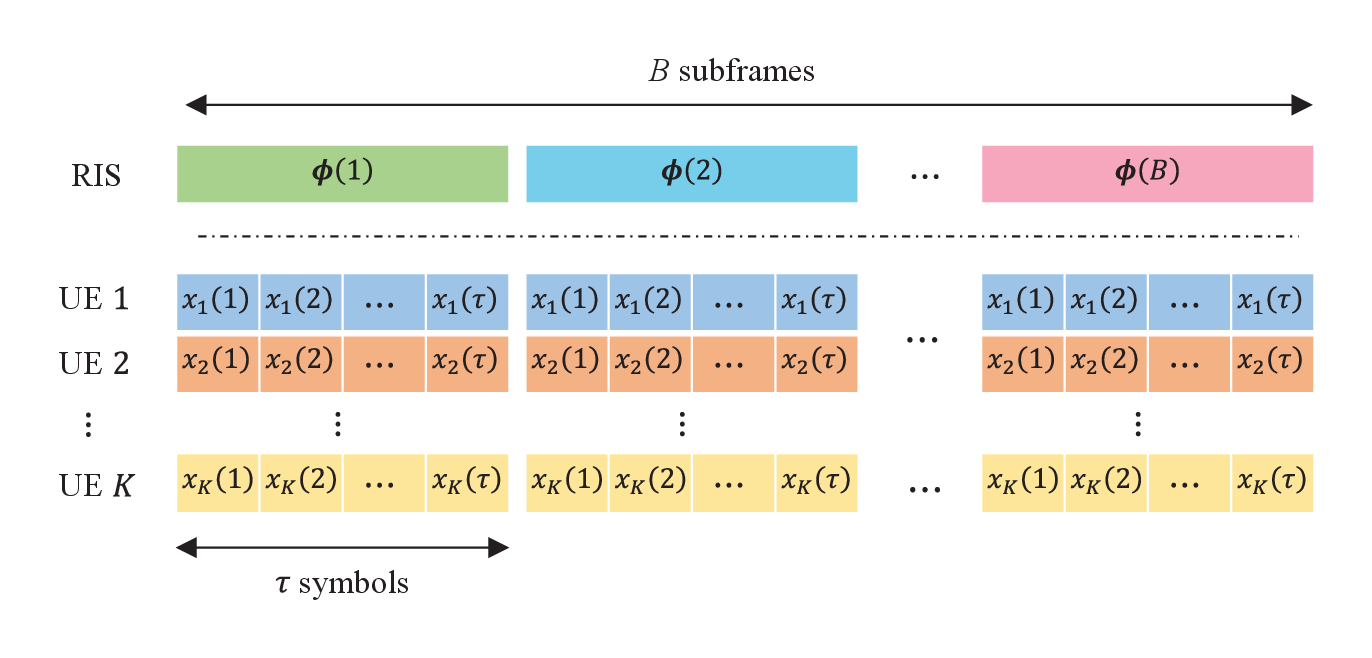}
\caption{The considered channel estimation protocol.}
\label{fig1:structure}
\end{figure*}

To characterize the RIS phase shifts, we utilize the parallel resonant circuit-based model considered in \cite{RISPracticalModel}. More specifically, the equivalent impedance of the $m$-th reflecting element, $m \in \mathcal M \triangleq \{1,\cdots,M\}$, is
expressed as
\begin{align}\label{Zm}
Z_m(C_m,R_m) =& \frac{j\omega L_1(j\omega L_2 + \frac{1}{j\omega C_m} + R_m)}{j\omega L_1 +( j\omega L_2 + \frac{1}{j\omega C_m} + R_m)},
\end{align}
where $C_m$, $R_m$, $L_1$, and $L_2$ stand for the effective capacitance, the effective resistance, the bottom layer inductance, and the top layer inductance of the parallel resonant circuit, respectively, and $\omega$ represents the angular frequency of the incident signal. Then, the reflection coefficient of the $m$-th element of the RIS is given by
\begin{align}\label{practicalmodel}
\phi_m = \beta_m e^{j \theta_m}=\frac{Z_m(C_m,R_m) - Z_0}{Z_m(C_m,R_m) + Z_0},
\end{align}
where $\beta_m \in [0,1]$ and $\theta_m \in [0,2\pi]$ stand for the reflection amplitude and the phase shift, respectively, and $Z_0$ denotes the free space impedance. Different from the commonly used unit modulus phase shift model, the amplitude and the phase of $\phi_m$ in (\ref{practicalmodel}) are coupled. Specifically, according to \cite{RISPracticalModel}, the reflection amplitude $\beta_m$ of the $m$-th element can be expressed as the following function of the corresponding phase shift $\theta_m$:
\begin{align}\label{fitness}
\beta_m (\theta_m) = (1-\beta_\text{min}) \left( \frac{\sin(\theta_m - \delta) + 1}{2} \right)^\alpha + \beta_\text{min},
\end{align}
where the values of the constants $\beta_\text{min} \geq 0$, $\delta \geq 0 $, and $\alpha \geq 0$ depend on the specific circuit parameters in (\ref{Zm}). Fig. \ref{fig:fittedcurve} illustrates an exemplified behaviour of the model in (\ref{fitness}), where $\alpha = 1.6$ and
$\delta = 0.43 \pi$.

\subsection{Cascaded Channel Estimation}
Let us focus on the channel estimation for the considered system.
An RIS cannot transmit training signals or perform channel estimation. Therefore, we cannot estimate $\mathbf H_r$ or $\mathbf G$ directly. Nevertheless, it is possible to estimate the cascaded channel of the reflected link at the BS.

We consider a block fading channel model where each block is divided into training-based channel estimation and data transmission stages. By adopting the channel estimation protocol in \cite{CE1,CEsparsity3}, as shown in Fig. \ref{fig1:structure} at the top of the next page, we further divide the training stage into $B$ subframes, each of which consists of $\tau$ symbols, where $\tau \geq K$ and $B \geq M+1$. The RIS coefficients are kept fixed within each subframe and vary for different subframes. The users transmit the same training signals periodically in different subframes.
Mathematically, in the $b$-th subframe, $b \in \mathcal B \triangleq \{1,\cdots,B\}$, the received signals of $\tau$ consecutive symbols at the BS, denoted by $\mathbf {Y}(b) \in \mathbb C^{L \times \tau}$, is given by
\begin{align}\label{signal_b}
\mathbf {Y}(b) =& \left( \mathbf{G} \mathbf{\Phi}(b) \mathbf{H}_r + \mathbf{H}_d\right) \mathbf{X} + \mathbf Z(b)\nonumber \\ =& \left(\sum_{m=1}^M \phi_m (b) \mathbf{g}_m \mathbf{h}_{r,m}^H  + \mathbf{H}_d\right) \mathbf{X} + \mathbf Z(b),
\end{align}
where $\mathbf{\Phi}(b) = \text{diag} \left\{[\phi_1(b), \cdots, \phi_M(b)]^T \right\}$ represents the fixed RIS reflection coefficients in subframe $b$ and $\mathbf{X}$ denotes the stacked training symbols of $K$ UEs, which is expressed as
\begin{align}\label{def:X}
\mathbf{X} \triangleq & \left[\mathbf x_1, \cdots, \mathbf x_K \right]^T \in \mathbb{C}^{K \times \tau}.
\end{align}
Here, $\mathbf x_k \triangleq [x_k(1), \cdots, x_k(\tau)]^T \in \mathbb{C}^{\tau \times 1}$ stands for the $\tau$ training symbols of the $k$-th UE, $k \in \mathcal K \triangleq \{1,\cdots, K\}$.
In addition, $\mathbf Z(b) \in \mathbb C^{L \times \tau}$ is the additive Gaussian white noise matrix in the $b$-th subframe, and $\mathbf{g}_m \in \mathbb{C}^{L \times 1}$ and $\mathbf{h}_{r,m} \in \mathbb{C}^{K \times 1}$ stand for the $m$-th column of $\mathbf{G}$ and $\mathbf{H}_r^H$, respectively.
To proceed, by setting $\phi_{M+1}(b) = 1$ and defining
\begin{align} \label{def:Gamma}
\mathbf{\Gamma}_m \triangleq &
 \begin{cases}
     \mathbf{g}_m \mathbf{h}_{r,m}^H  &m \in \mathcal M  \\
    \mathbf{H}_d   &m = M+1,
 \end{cases}
\end{align}
the received signal $\mathbf Y(b)$ in (\ref{signal_b}) becomes
\begin{align}\label{signal_n}
\mathbf Y(b) =&\  \left(\sum_{m=1}^{M+1} \phi_m(b) \mathbf{\Gamma}_m  \right)\mathbf X + \mathbf Z(b) \nonumber \\
\triangleq &\ [\mathbf{\Gamma}_1, \cdots , \mathbf{\Gamma}_{M+1}] ( \boldsymbol \phi(b) \otimes \mathbf I_K ) \mathbf X + \mathbf Z(b) \nonumber \\
\triangleq &\ \mathbf{\Gamma} ( \boldsymbol \phi(b) \otimes \mathbf I_K ) \mathbf X + \mathbf Z(b),
\end{align}
where $\boldsymbol \phi(b) \triangleq [\phi_1(b),\cdots , \phi_{M+1}(b)]^T$ and $\mathbf{\Gamma} \triangleq [\mathbf{\Gamma}_1, \cdots , \mathbf{\Gamma}_{M+1}]$ is the equivalent channel to be estimated in this work, which consists of the direct and cascaded reflected channels.
Furthermore, by gathering the received signal $\mathbf Y(b)$ of all the $B$ subframes, we obtain
\begin{align}\label{receivesignalN}
\mathbf Y = \mathbf \Gamma \mathbf S + \mathbf Z,
\end{align}
where
$\mathbf Y  \triangleq \left[\mathbf Y(1), \cdots, \mathbf Y(B)\right] \in \mathbb{C}^{L \times \tau B}$ and $\mathbf Z  \triangleq \left[\mathbf Z(1), \cdots, \mathbf Z(B)\right] \in \mathbb{C}^{L \times \tau B}$
denote the received signals and the noises of the whole training phase, respectively, and $\mathbf S$ is relevant to both the training sequence and the reflection pattern, which is expressed as
\begin{align}\label{def:S}
\mathbf S  &\triangleq [ ( \boldsymbol \phi(1) \otimes \mathbf I_K )\mathbf{X}, \cdots,  ( \boldsymbol \phi(B) \otimes \mathbf I_K ) \mathbf{X}] \in \mathbb{C}^{[(M+1)K] \times \tau B}.
\end{align}
Define the reflection pattern of $B$ subframes at the RIS by
\begin{align}\label{def:V}
\mathbf{V} \triangleq & \left[\boldsymbol \phi(1), \cdots, \boldsymbol \phi(B)\right] \in \mathbb{C}^{(M+1) \times B},
\end{align}
each column of which represents the reflection pattern of the corresponding subframe. Then, based on the definitions in (\ref{def:S}) and (\ref{def:V}), we obtain the following expression for $\mathbf S$
\begin{align} \label{def:widetilde}
\mathbf S = (\mathbf{V} \otimes \mathbf I_K)  (\mathbf I_B \otimes \mathbf{X})
\triangleq \mathbf{\widetilde V} \mathbf{\widetilde X},
\end{align}
where $\mathbf{\widetilde V} \triangleq \mathbf{V} \otimes \mathbf I_K$ and $\mathbf{\widetilde X} \triangleq \mathbf I_B \otimes \mathbf{X}$ are of size $[(M+1)K] \times BK$ and $KB \times \tau B$, respectively.

The goal of channel estimation for the considered RIS-aided multiuser system is to recover the channel matrix $\mathbf \Gamma$ based on the knowledge of the received signal matrix $\mathbf{Y}$ and a properly designed matrix $\mathbf S$.
From (\ref{receivesignalN}), we obtain the LS and the LMMSE estimates of $\mathbf \Gamma$ as follows \cite{MIMOChannelEstimation}:
\begin{align}
\mathbf{\hat \Gamma}_\text{LS}&= \mathbf{Y} \mathbf S^\dag, \label{hatGLS}\\
\mathbf{ \hat \Gamma}_\text{LMMSE} &= \mathbf Y \left(\mathbf S^H \mathbf R_{\mathbf \Gamma}  \mathbf S + \sigma^2 L \mathbf I_{\tau B}\right)^{-1}\mathbf S^H \mathbf R_{\mathbf \Gamma} \label{hatGLMMSE},
\end{align}
where $\mathbf S^\dag = \mathbf S^H (\mathbf S \mathbf S^H)^{-1}$ denotes the pseudoinverse of $\mathbf S$, $\mathbf R_{\mathbf \Gamma} = \mathbb{E} \{ \mathbf \Gamma^H \mathbf \Gamma\}$ is the correlation matrix of $\mathbf \Gamma$, and $\sigma^2$ denotes the noise variance.
The estimation errors corresponding to (\ref{hatGLS}) and (\ref{hatGLMMSE}) are given by
\begin{align}
J_\text{LS} = &\
\mathbb{E}\left\{ \| \mathbf{ \hat \Gamma}_\text{LS} - \mathbf \Gamma\|_F^2 \right\}
=  \sigma^2 L \text{Tr}\left[\left( \mathbf S \mathbf S^H \right)^{-1}\right],\label{LSMSE}\\
J_\text{LMMSE} = &\
 \mathbb{E}\left\{\| \mathbf{ \hat \Gamma}_\text{LMMSE} - \mathbf \Gamma\|_F^2  \right\} \nonumber\\
=&\  \text{Tr}\left[\left(\mathbf R_{\mathbf \Gamma}^{-1} + \frac{1}{\sigma^2 L} \mathbf S \mathbf S^H \right)^{-1}\right].\label{MMSEMSE}
\end{align}

In addition, the training overhead of the considered estimation scheme is $(M + 1)K$. There exist some possible approaches to alleviate this overhead, e.g., by grouping the RIS elements \cite{onoffgroup} and by adopting the two-timescale protocol \cite{twotimescale1,twotimescale2}, which will be discussed in Section VI-C.

\subsection{Problem Formulation}
In this paper, we aim to design the matrix $\mathbf S$, i.e., the training sequence $\mathbf{X}$ at the UE side and the reflection coefficient $\mathbf V$ at the non-ideal RIS, such that the channel estimation MSEs $J_\text{LS}$ or $J_\text{LMMSE}$ are minimized. The considered joint optimization problem is formulated as
\begin{align}\label{iniProb}
\mathop \text{minimize}\limits_{\mathbf X, \mathbf{V}} \quad & J(\mathbf{X},\mathbf{V}) \nonumber \\
\text{subject to} \quad & \|\mathbf x_k\|^2 \leq P_k,\ k \in \mathcal K \nonumber \\
& [\mathbf V]_{m,n} \in \mathcal{F}, \ m \in \mathcal M, \ n \in \mathcal B\nonumber\\
& [\mathbf V]_{M+1,n} = 1, \ n \in \mathcal B,
\end{align}
where $J(\mathbf{X},\mathbf{V})$ is given by (\ref{LSMSE}) for the LS estimator and by (\ref{MMSEMSE}) for the LMMSE estimator. $\|\mathbf x_k\|^2 \leq P_k$ stands for the transmit power constraint of UE $k$ and $P_k$ is the corresponding power budget. $[\mathbf{V}]_{m,n} \in \mathcal{F}$ means that the entry $[\mathbf{V}]_{m,n}$ satisfies the phase shift constraint given in (\ref{fitness}). $[\mathbf V]_{M+1,n} = 1 $ follows from the definition $\phi_{M+1}(b) = 1$.

Solving the problem in (\ref{iniProb}) is difficult due to the coupled variables $\mathbf X$ and $\mathbf V$ in the objective function and the realistic constraint for the reflection coefficient. We develop efficient algorithms to tackle the problems for the LS and LMMSE channel estimators in the following two sections, respectively.

\begin{remark}
When considering multi-antenna UEs, we define $\mathbf X$ as $[\mathbf X_1^T, \cdots, \mathbf X_K^T ]^T \in \mathbb C^{N \times \tau}$, where $\mathbf X_k \in \mathbb C^{N_k \times \tau}$ denotes the pilot matrix sent by the $k$-th UE equipped with $N_k$ antennas and $\tau \geq N \triangleq \sum_{k=1}^K N_k$. The optimization problem can be formulated similarly, where the dimensions of $\mathbf H_r$, $\mathbf H_d$, $\mathbf \Gamma$, and $\mathbf S$ need to be adjusted, and the methods proposed in the following sections can be modified and applied accordingly.
\end{remark}

\section{LS Channel Estimation}
In this section, we focus on the optimization problem in (\ref{iniProb}) for the LS channel estimator.
Firstly, we prove that the optimal training matrix is independent of the reflection pattern and has an interesting orthogonal property.
Then, we propose an efficient MM-based algorithm to optimize the RIS reflection pattern under the considered circuit model for the RIS element, where a semi-closed form solution is derived in each iteration.

\subsection{Closed-Form Optimal Training Matrix}
Although the training matrix and the reflection pattern are coupled in the objective function $J_\text{LS}$ in problem (\ref{iniProb}), as stated in the following theorem, we can determine the optimal training matrix in a closed form for the LS channel estimator.
\begin{theorem}\label{theorem1}
The optimal training matrix for the LS channel estimator of the considered non-ideal RIS-assisted system satisfies the condition:
\begin{align}\label{optimalX}
\mathbf X \mathbf X^H = \text{diag} \{ P_1, \cdots, P_K\}.
\end{align}
\end{theorem}
\begin{IEEEproof}
See Appendix \ref{proof:Theorem1}.
\end{IEEEproof}

It follows from Theorem \ref{theorem1} that any orthogonal training matrix is optimal for the LS channel estimator, which conforms to the results in MIMO communications\cite{MIMOChannelEstimation} and ideal RIS-assisted communication systems\cite{fullpattern}. Herein, we adopt a simple discrete Fourier transform (DFT)-based training matrix for estimating the uplink multiuser channel. Specifically, the training sequence of each UE is expressed as
\begin{align}\label{LSoptX}
\mathbf x_k = \frac{\sqrt{P_k}}{\| \mathbf d_k \|} \mathbf d_k, \ k \in \mathcal K,
\end{align}
where $\mathbf d_k$ denotes the $k$-th column of the $\tau \times \tau$ DFT matrix.

\subsection{Optimization of RIS Reflection Pattern}
In this subsection, we consider the optimization with respect to the RIS reflection pattern under the considered non-ideal phase shift constraint, which, according to the proof of Theorem \ref{theorem1}, takes the form:
\begin{align}\label{prob:LSV1}
\mathop \text{minimize}\limits_{ \mathbf{V}} \quad &  \text{Tr}\left[\left( \mathbf{V}  \mathbf{ V}^H \right)^{-1}\right] \nonumber \\
\text{subject to} \quad
& [\mathbf V]_{m,n} \in \mathcal{F}, \ m \in \mathcal M, \ n \in \mathcal B\nonumber\\
& [\mathbf V]_{M+1,n} = 1, \ n \in \mathcal B.
\end{align}
Although the variable $\mathbf{X}$ has been eliminated, it is still nontrivial to find the optimal solution of problem (\ref{prob:LSV1}) mainly due to the intricate phase shift constraint. In fact, although it has been proved in \cite{fullpattern} that the optimal reflection pattern minimizing the MSE of the LS channel estimator is an orthogonal matrix, e.g., the DFT matrix or the Hadamard matrix, for ideal unit-modulus RIS phase shifts, this conclusion does not necessarily hold for the non-ideal phase shift constraints considered in this work. Hence, to design the non-ideal RIS reflection pattern for the LS channel estimator, we develop an MM-based algorithm.

The basic idea of the MM algorithm is to find a series of simple surrogate problems whose objective functions are locally approximated to the original objective function in each iteration, and then iteratively solve the surrogate problems until convergence \cite{PalomarMMAlg}.
To do this, we first derive an appropriate surrogate function with a more tractable form to locally approximate the objective function of problem (\ref{prob:LSV1}), as given in the following proposition.

\begin{proposition}\label{LSlemma1}
For the objective function $\text{Tr}[( \mathbf{V} \mathbf{V}^H)^{-1}]$, a surrogate upper bound for the $i$-th iteration is
\begin{align}\label{up:LSV}
 f\left( \mathbf{V}; \mathbf{V}_0 \right)
= \lambda_1 \text{Tr}\left[ \mathbf{V} \mathbf{V}^H \right] + 2\mathcal R\left\{ \text{Tr}\left[ \mathbf{A}_0 \mathbf{V} \right]\right\} +\text{Const}(\mathbf{V}_0),
\end{align}
where $\mathbf{ V}_0$ denotes the solution to $\mathbf{V}$ in the $(i-1)$-th iteration of the MM algorithm, and
$\lambda_1$, $\mathbf{A}_0$, and $\text{Const}(\mathbf{V}_0)$ are defined as follows:
\begin{align}
\lambda_1 \triangleq &\ 3 \text{Tr}\left[\left( \mathbf{V}_0 \mathbf{V}_0^H \right)^{-1}\right]^2,  \nonumber \\
\mathbf{A}_0 \triangleq &\ -  \mathbf{ V}_0^H  \left(\mathbf{ V}_0 \mathbf{ V}_0^H\right)^{-2}- \lambda_1   \mathbf{ V}_0^H,  \nonumber \\
\text{Const}(\mathbf{V}_0) = &\ \text{Tr}\left[\left( \mathbf{ V}_0 \mathbf{ V}_0^H \right)^{-1}\right] + \lambda_1 \text{Tr}\left[ \mathbf{V}_0 \mathbf{V}_0^H \right] \nonumber \\
&\  +  2 \text{Tr}\left[  \mathbf{ V}_0^H  \left(\mathbf{ V}_0 \mathbf{ V}_0^H\right)^{-2} \mathbf{ V}_0 \right].
\end{align}
\end{proposition}
\begin{IEEEproof}
See Appendix~\ref{proof1}.
\end{IEEEproof}

The major advantage of constructing the surrogate function in (\ref{up:LSV}) is that the intractable inversion operation in the objective function of (\ref{prob:LSV1}) is removed and we can perform the minimization of (\ref{up:LSV}) by optimizing each entry of $\mathbf{V}$ simultaneously. Specifically, replacing the objective function in (\ref{prob:LSV1}) with $ f\left( \mathbf{V}; \mathbf{V}_0 \right)$, we obtain the following problem
\begin{align}\label{prob:LSV2}
\mathop \text{minimize}\limits_{ \mathbf{V}} \quad &  \lambda_1 \text{Tr}\left[ \mathbf{V} \mathbf{V}^H \right] + 2 \mathcal R\left\{ \text{Tr}\left[ \mathbf{A}_0 \mathbf{V} \right]\right\} \nonumber \\
\text{subject to} \quad
& [\mathbf V]_{m,n} \in \mathcal{F}, \ m \in \mathcal M, \ n \in \mathcal B\nonumber\\
& [\mathbf V]_{M+1,n} = 1, \ n \in \mathcal B.
\end{align}
In particular, we show that the optimal solution to problem (\ref{prob:LSV2}) can be determined in an element-wise manner.

\begin{proposition}\label{prop:LSexhaustive}
The element-wise optimal solution of problem (\ref{prob:LSV2}) in the $i$-th iteration is obtained by
\begin{align}\label{LSThetaUpdateRule}
\left[\mathbf{V}\right]_{m,n} =
 \begin{cases}
    \beta(\bar\theta_{m,n}) e^{j \bar\theta_{m,n}}   \ &m \in \mathcal M, \ n \in \mathcal B \\
   \quad 1   &m = M+1,
 \end{cases}
\end{align}
where $\bar\theta_{m,n} = \mathop{\arg\min}\limits_{\theta \in [0,2\pi] }
\bar f_{m,n}(\theta)$ and $\bar f_{m,n}(\theta)$ is given by
\begin{align}\label{def:f_mn_theta}
&\ \bar f_{m,n}(\theta) \nonumber\\
\!=&\ \lambda_1\! \left( \xi^2(\sin(\theta\!-\!\delta)\!+\!1)^{2\alpha} \!+\! \beta_\text{min}^2 \!+\! 2\xi \beta_\text{min} (\sin(\theta\!-\!\delta)\!+\!1)^\alpha   \right) \nonumber\\
&\! +\!\! 2|[\mathbf{A}_0]_{n,m}| \! (\xi(\sin(\theta\!-\!\delta)\!+\!1)^\alpha \!\!+\! \beta_\text{min}) \cos(\arg([\mathbf{A}_0]_{n,m} ) \!+\! \theta)
\end{align}
with $\xi \triangleq (1 - \beta_\text{min})\left(\frac{1}{2}\right)^\alpha$.
The minimization of $ \bar f_{m,n}(\theta)$ can be obtained by performing a one-dimensional search over $\theta \in [0,2\pi]$.
\end{proposition}
\begin{IEEEproof}
See Appendix \ref{proof:LSexhaustive}.
\end{IEEEproof}

With the solution of the RIS reflection pattern given in (\ref{LSThetaUpdateRule}), the proposed MM-based algorithm for the LS channel estimator is summarized in Algorithm \ref{alg:LSAO}. Moreover, we show the convergence of the MM algorithm in the following theorem.

\begin{theorem}\label{theorem2}
The proposed MM algorithm always converges to a stationary point of the problem in (\ref{prob:LSV1}).
\end{theorem}
\begin{IEEEproof}
See Appendix~\ref{proof:theorem2}.
\end{IEEEproof}

\begin{remark}
In Proposition \ref{prop:LSexhaustive}, we address the element-wise optimization of $[\mathbf V]_{m,n}$ by performing the one-dimensional search over $\theta \in [0,2\pi]$ based on the equivalent model in (\ref{fitness}). In fact, the reflection coefficient can be modeled by the formulation in (\ref{practicalmodel}) directly, and the optimization of $[\mathbf V]_{m,n}$ can also be obtained by searching for the effective capacitance $C_m$.
\end{remark}

\begin{algorithm}[t]
\caption{Proposed Algorithm for LS Channel Estimator}
\label{alg:LSAO}
\begin{algorithmic}[1]
\STATE \textit{Initialization:} Set the initial point $\mathbf{V}^{(0)}$, iteration index $i = 0$, and convergence accuracy $\epsilon$.
\STATE Obtain the optimal $\mathbf{X}$ according to (\ref{LSoptX}).
\REPEAT
\STATE Set $i=i+1$.\\

    \STATE
    Calculate $\mathbf{V}^{(i)}$ according to (\ref{LSThetaUpdateRule});
\UNTIL convergence.
\STATE Output the optimal $\mathbf{V}^{(i)}$.
\end{algorithmic}
\end{algorithm}

\section{LMMSE Channel Estimation}
In this section, we address the optimization problem in (\ref{iniProb}) for the LMMSE channel estimator, which can be applied if the channel correlation matrix is known. To begin with, we utilize the matrix inversion lemma \cite{matrixinversionlemma} and transform $J_\text{LMMSE}$ as follows:
\begin{align}
 J_\text{LMMSE}
= &\ \text{Tr}\left[\left(\mathbf R_{\mathbf \Gamma}^{-1} + \frac{1}{\sigma^2 L} \mathbf S \mathbf S^H \right)^{-1}\right] \nonumber\\
= &\ \text{Tr}\left[\mathbf R_{\mathbf \Gamma} - \mathbf R_{\mathbf \Gamma} \mathbf S \left( \mathbf S^H \mathbf R_{\mathbf \Gamma} \mathbf S + \sigma^2 L \mathbf I_{\tau B} \right)^{-1} \mathbf S^H \mathbf R_{\mathbf \Gamma}\right]\label{MSEnew}.
\end{align}
As a result, the corresponding MSE minimization problem is equivalently reformulated as
\begin{align}\label{LMMSEiniProb}
\mathop \text{minimize}\limits_{\mathbf X, \mathbf{V}} \quad &
- \text{Tr}\left[  \mathbf R_{\mathbf \Gamma} \mathbf S \left( \mathbf S^H \mathbf R_{\mathbf \Gamma} \mathbf S + \sigma^2 L \mathbf I_{\tau B}\right)^{-1} \mathbf S^H \mathbf R_{\mathbf \Gamma}\right] \nonumber \\
\text{subject to} \quad & \|\mathbf x_k\|^2 \leq P_k,\ k  \in \mathcal K \nonumber \\
& [\mathbf V]_{m,n} \in \mathcal{F}, \ m \in \mathcal M, \ n \in \mathcal B \nonumber\\
& [\mathbf V]_{M+1,n} = 1, \ n \in \mathcal B.
\end{align}
Compared to the LS channel estimator, the objective function of problem (\ref{LMMSEiniProb}), denoted as $g(\mathbf S)$, is more complex, thus making it harder to optimize $\mathbf X$ and $\mathbf V$ separately. Hence, in order to address the considered problem, we propose an MM-based alternating algorithm.

We first construct a tractable surrogate function, as shown in the subsequent lemma, to iteratively upper bound $g(\mathbf S)$.
\begin{lemma}\label{lemma:LMMSElemma1}
For problem (\ref{LMMSEiniProb}), $g(\mathbf S)$ is upper bounded by $g(\mathbf S;\mathbf S_0)$ given as follows:
\begin{align}\label{JMMSEub1}
 \ g(\mathbf S;\mathbf S_0) \triangleq&\  \text{Tr}\left[\mathbf \Xi_0^H \mathbf S^H \mathbf R_{\mathbf \Gamma} \mathbf S \mathbf \Xi_0  \right]
 -  2\mathcal R\left\{\text{Tr}\left[ \mathbf \Xi_0 \mathbf R_{\mathbf \Gamma} \mathbf S\right]\right\} \nonumber\\
&\ + \sigma^2 L \text{Tr}\left[ \mathbf \Xi_0 \mathbf \Xi_0^H  \right],
\end{align}
where $\mathbf \Xi_0 = \left(\mathbf S_0^H \mathbf R_{\mathbf \Gamma} \mathbf S_0 + \sigma^2 L \mathbf I_{\tau B}\right)^{-1}\mathbf S_0^H \mathbf R_{\mathbf \Gamma}$
and $\mathbf S_0$ is an arbitrary feasible solution to $\mathbf S$.
\end{lemma}
\begin{IEEEproof}
See Appendix \ref{proof:concave}.
\end{IEEEproof}
We observe that $g(\mathbf S;\mathbf S_0)$ in (\ref{JMMSEub1}) only consists of a quadratic term and a linear term with respect to $\mathbf S$, which makes it much easier to be handled than the original $g(\mathbf S)$. Based on Lemma~\ref{lemma:LMMSElemma1}, we address the minimization of $g(\mathbf S)$ by iteratively minimizing $g(\mathbf S;\mathbf S_0)$, where the variables $\mathbf X$ and $\mathbf V$ are optimized in an alternating manner.

\subsection{Optimization of the Training Matrix}
We first perform the optimization of $\mathbf X$ with a fixed $\mathbf V$. Substituting $\mathbf S = \mathbf{\widetilde{V}} \mathbf{\widetilde{X}}$ into $g(\mathbf S;\mathbf S_0)$ and omitting the constant terms, the training sequence design under the per-UE transmitter power constraints is formulated as
\begin{align}\label{prob:LMMSEX1}
\mathop \text{minimize}\limits_{ \mathbf{X}} \quad &
\text{Tr}\left[\mathbf \Xi_0^H \mathbf{\widetilde{X}}^H\mathbf{\widetilde{V}}_0^H \mathbf R_{\mathbf \Gamma} \mathbf{\widetilde{V}}_0 \mathbf{\widetilde{X}} \mathbf \Xi_0  \right] \nonumber \\
& - 2\mathcal R\left\{\text{Tr}\left[  \mathbf \Xi_0 \mathbf R_{\mathbf \Gamma} \mathbf{\widetilde{V}}_0 \mathbf{\widetilde{X}}\right]\right\} \nonumber \\
\text{subject to} \quad & \|\mathbf x_k\|^2 \leq P_k, \ k \in \mathcal K,
\end{align}
where $\mathbf{\widetilde{V}}_0 = \mathbf{V}_0 \otimes \mathbf I_K$ and $\mathbf{V}_0$ represents the fixed reflection pattern.
Problem (\ref{prob:LMMSEX1}) can be formulated into a convex problem with respect to the block-diagonal matrix $\mathbf {\widetilde{X}}$, as follows:
\begin{align}\label{prob:widetildeX}
\mathop \text{minimize}\limits_{ \mathbf {\widetilde{X}}} \quad &
\text{Tr}\left[\mathbf \Xi_0^H \mathbf{\widetilde{X}}^H\mathbf{\widetilde{V}}_0^H \mathbf R_{\mathbf \Gamma} \mathbf{\widetilde{V}}_0 \mathbf{\widetilde{X}} \mathbf \Xi_0  \right] \nonumber \\
& - 2\mathcal R\left\{\text{Tr}\left[  \mathbf \Xi_0 \mathbf R_{\mathbf \Gamma} \mathbf{\widetilde{V}}_0 \mathbf{\widetilde{X}}\right]\right\} \nonumber \\
\text{subject to} \quad & \mathbf {\widetilde{X}} = \mathbf I_B \otimes \mathbf{X}, \nonumber \\
&  \|[\mathbf{X}]_{k,1:\tau}^T\|^2 \leq P_k, \ k \in \mathcal K,
\end{align}
and then solved. However, directly solving (\ref{prob:widetildeX}) can be computationally inefficient due to the large dimension of $\mathbf {\widetilde{X}}$.
Therefore, we propose an MM-based solution for problem (\ref{prob:LMMSEX1}), which only requires calculating closed-form expressions iteratively and results in a lower computational complexity.

Specifically, we employ the following upper bound to the first term in the objective function, according to \cite[eq. (26)]{PalomarMMAlg}:
\begin{align}\label{up:LMMSEX1}
& \text{Tr}\left[\mathbf \Xi_0^H \mathbf{\widetilde{X}}^H\mathbf{\widetilde{V}}_0^H \mathbf R_{\mathbf \Gamma} \mathbf{\widetilde{V}}_0 \mathbf{\widetilde{X}} \mathbf \Xi_0  \right] \nonumber\\
= &\ \text{vec}^H( \mathbf{\widetilde{X}})
\left( \left( \mathbf \Xi_0 \mathbf \Xi_0^H \right)^T \otimes \mathbf{\widetilde{V}}_0^H \mathbf R_{\mathbf \Gamma} \mathbf{\widetilde{V}}_0  \right)
\text{vec}(\mathbf{\widetilde{X}}) \nonumber\\
\leq &\ \lambda_2 \| \mathbf{\widetilde{X}}\|_F^2 - 2 \mathcal R \left\{ \text{Tr} \left[\lambda_2 \mathbf{\widetilde{X}}_0^H \mathbf{\widetilde{X}}
-  \mathbf \Xi_0 \mathbf \Xi_0^H \mathbf{\widetilde{X}}_0^H \mathbf{\widetilde{V}}_0^H \mathbf R_{\mathbf \Gamma} \mathbf{\widetilde{V}}_0 \mathbf{\widetilde{X}} \right] \right\} \nonumber\\
&\ + \text{vec}^H(\mathbf{\widetilde{X}}_0) \left( \lambda_2 \mathbf I - \left( \mathbf \Xi_0 \mathbf \Xi_0^H \right)^T \otimes \mathbf{\widetilde{V}}_0^H \mathbf R_{\mathbf \Gamma} \mathbf{\widetilde{V}}_0 \right)    \text{vec}(\mathbf{\widetilde{X}}_0),
\end{align}
where $\lambda_2 \mathbf I \succeq  ( \mathbf \Xi_0 \mathbf \Xi_0^H )^T \otimes \mathbf{\widetilde{V}}^H_0 \mathbf R_{\mathbf \Gamma} \mathbf{\widetilde{V}}_0$ and $\mathbf{\widetilde{X}}_0$ is the solution of $\mathbf{\widetilde{X}}$ in the previous iteration.
For simplicity, the value of $\lambda_2$ can be set to $\lambda_2 =\lambda_\text{max} ( ( \mathbf \Xi_0 \mathbf \Xi_0^H )^T \otimes \mathbf{\widetilde{V}}^H_0 \mathbf R_{\mathbf \Gamma} \mathbf{\widetilde{V}}_0  )= \lambda_\text{max} ( \mathbf \Xi_0 \mathbf \Xi_0^H ) \lambda_\text{max} (\mathbf{\widetilde{V}}_0^H \mathbf R_{\mathbf \Gamma} \mathbf{\widetilde{V}}_0)$.

By substituting (\ref{up:LMMSEX1}) and removing the constant terms, we transform problem (\ref{prob:LMMSEX1}) into
\begin{align}\label{prob:LMMSEX2}
\mathop \text{minimize}\limits_{ \mathbf{X}} \quad &
\lambda_2 \text{Tr}\left[\mathbf{\widetilde{X}} \mathbf{\widetilde{X}}^H \right]
- 2\mathcal R\left\{\text{Tr}\left[  \mathbf{B}_0 \mathbf{\widetilde{X}}\right]\right\} \nonumber \\
\text{subject to} \quad & \|\mathbf x_k\|^2 \leq P_k,\ k \in \mathcal K,
\end{align}
where $
\mathbf{B}_0
= \lambda_2 \mathbf{\widetilde X}_0^H -
\mathbf \Xi_0\mathbf \Xi_0^H \mathbf{\widetilde X}_0^H \mathbf{\widetilde V}_0^H \mathbf R_{\mathbf \Gamma} \mathbf{\widetilde V}_0+ \mathbf \Xi_0 \mathbf R_{\mathbf \Gamma} \mathbf{\widetilde V}_0$.
By invoking the definition $\mathbf{\widetilde{X}} = \mathbf I_B \otimes \mathbf{X}$ in (\ref{def:X}), we can readily decompose the problem in (\ref{prob:LMMSEX2}) into $K$ subproblems, with respect to the training sequence design at each UE. Concretely, each subproblem is given by
\begin{align}\label{prob:LMMSperUE}
\mathop \text{minimize}\limits_{\mathbf{x}_k} \quad &   \lambda_2 B \| \mathbf x_k \|^2 -  2\mathcal R\left\{ \text{Tr}\left[ \mathbf{b}^H_k  \mathbf x_k  \right]\right\} \nonumber \\
\text{subject to} \quad & \| \mathbf x_k \|^2 \leq P_k,
\end{align}
where $\mathbf{b}_k \triangleq \sum_{b=1}^B \left[ \mathbf{B}_0 \right]^*_{(b-1)\tau+1:b\tau,(b-1)K+k}$.
Note that (\ref{prob:LMMSperUE}) is a convex quadratic optimization problem which can be readily tackled. In fact,
there exists a closed-form optimal solution as shown in the following proposition.

\begin{proposition}\label{prop:LMMSEX}
The optimal solution to problem (\ref{prob:LMMSperUE}) can be expressed in a closed form as follows:
\begin{align}\label{LMMSEXupdate}
\mathbf{x}_k =
\begin{cases}
\frac{\sqrt{P_k}}{\| \mathbf b_k \|} \mathbf b_k \quad &\text{if} \quad \| \mathbf b_k \| >  \sqrt{P_k}\lambda_2 B, \\
 \quad \frac{\mathbf b_k}{ \lambda_2 B } \quad &\text{if} \quad \| \mathbf b_k \| \leq \sqrt{P_k}\lambda_2 B.
\end{cases}
\end{align}
\end{proposition}
\begin{IEEEproof}
See Appendix~\ref{proof:KKT1}.
\end{IEEEproof}
By invoking (\ref{LMMSEXupdate}) for all the $K$ UEs, we accomplish the optimization of $\mathbf X$ with a low computational cost.

\subsection{Optimization of the RIS Reflection Pattern}
Now we fix $\mathbf X$ and focus on the optimization of the RIS reflection pattern. According to the upper bound $g(\mathbf S;\mathbf S_0)$ in (\ref{JMMSEub1}), we formulate the subproblem with respect to $\mathbf V$ as
\begin{align}\label{prob:LMMSEV1}
\mathop \text{minimize}\limits_{  \mathbf{V}} \quad &
\text{Tr}\left[\mathbf \Xi_0^H \mathbf{\widetilde{X}}_0^H\mathbf{\widetilde{V}}^H \mathbf R_{\mathbf \Gamma} \mathbf{\widetilde{V}} \mathbf{\widetilde{X}}_0 \mathbf \Xi_0  \right] \nonumber\\
& - 2\mathcal R\left\{\text{Tr}\left[ \mathbf{\widetilde{X}}_0 \mathbf \Xi_0 \mathbf R_{\mathbf \Gamma} \mathbf{\widetilde{V}}\right]\right\} \nonumber \\
\text{subject to} \quad
& [\mathbf V]_{m,n} \in \mathcal{F}, \ m \in \mathcal M, \ n \in \mathcal B \nonumber\\
& [\mathbf V]_{M+1,n} = 1, \ n \in \mathcal B.
\end{align}
Similar to the optimization of the training matrix tackled in the previous subsection, by employing the upper bound given in \cite[eq. (26)]{PalomarMMAlg}, we obtain
\begin{align}\label{up:LMMSEV1}
&\text{Tr}\left[\mathbf \Xi_0^H \mathbf{\widetilde{X}}_0^H\mathbf{\widetilde{V}}^H \mathbf R_{\mathbf \Gamma} \mathbf{\widetilde{V}} \mathbf{\widetilde{X}}_0 \mathbf \Xi_0  \right] \nonumber\\
\leq & \lambda_3 \|\mathbf{\widetilde{V}}\|_F^2 -2\mathcal R\left\{\text{Tr}\left[\lambda_3 \mathbf{\widetilde{V}}_0^H \mathbf{\widetilde{V}} - \mathbf{\widetilde{X}}_0 \mathbf \Xi_0\mathbf \Xi_0^H  \mathbf{\widetilde{X}}_0^H \mathbf{\widetilde{V}}_0^H  \mathbf R_{\mathbf \Gamma}  \mathbf{\widetilde{V}}\right]\right\} \nonumber\\
& + \text{vec}^H(\mathbf{\widetilde{V}}_0)
\left( \lambda_3 \mathbf I -  \left( \mathbf{\widetilde{X}}_0 \mathbf \Xi_0 \mathbf \Xi_0^H \mathbf{\widetilde{X}}_0^H \right)^T \otimes \mathbf R_{\mathbf \Gamma}  \right)
\text{vec}(\mathbf{\widetilde{V}}_0),
\end{align}
and transform problem (\ref{prob:LMMSEV1}) into
\begin{align}\label{prob:LMMSEV2}
\mathop \text{minimize}\limits_{ \mathbf{V}} \quad &
\lambda_3 \text{Tr} \left[  \mathbf{\widetilde{V}}  \mathbf{\widetilde{V}}^H \right]
- 2\mathcal R\left\{\text{Tr}\left[ \mathbf{C}_0 \mathbf{\widetilde{V}}\right]\right\} \nonumber \\
\text{subject to} \quad
& [\mathbf V]_{m,n} \in \mathcal{F}, \ m \in \mathcal M, \ n \in \mathcal B\nonumber\\
& [\mathbf V]_{M+1,n} = 1, \ n \in \mathcal B,
\end{align}
where $\mathbf{C}_0 = \lambda_3  \mathbf{\widetilde{V}}_0^H
- \mathbf{\widetilde{X}}_0 \mathbf \Xi_0\mathbf \Xi_0^H  \mathbf{\widetilde{X}}_0^H \mathbf{\widetilde{V}}_0^H  \mathbf R_{\mathbf \Gamma}
+ \mathbf{\widetilde{X}}_0 \mathbf \Xi_0\mathbf R_{\mathbf \Gamma}$, $\mathbf{\widetilde{V}}_0$ is the solution of $\mathbf{\widetilde{V}}$ in the previous iteration, and $\lambda_3$ is calculated as $ \lambda_\text{max} (\mathbf{\widetilde{X}}_0 \mathbf \Xi_0 \mathbf \Xi_0^H \mathbf{\widetilde{X}}_0^H )\lambda_\text{max} \left(  \mathbf R_{\mathbf \Gamma}\right)$.

By substituting $\mathbf{\widetilde{V}} \triangleq \mathbf{V} \otimes \mathbf I_K$ into the objective function of the problem in (\ref{prob:LMMSEV2}), we obtain $\sum_m \sum_n (\lambda_3 K |[\mathbf{V}]_{m,n}|^2 -2\mathcal R\left\{ c_{m,n}[\mathbf{V}]_{m,n} \right\})$, where $c_{m,n} = \sum_{k=1}^K \left[\mathbf{C}_0\right]_{(n-1)K+k,(m-1)K+k}$. Hence, we can solve problem (\ref{prob:LMMSEV2}) in an element-wise manner. In particular, the solution to each entry
of $\mathbf{V}$ in the $i$-th iteration can be obtained according to
\begin{align}\label{LMMSEV}
\left[\mathbf{V}\right]_{m,n} =
 \begin{cases}
    \beta(\bar\theta_{m,n}) e^{j \bar\theta_{m,n}}  \ &m \in \mathcal M, \ n \in \mathcal B \\
   \quad 1   &m = M+1,
 \end{cases}
\end{align}
where $ \bar\theta_{m,n} = \mathop{\arg\min}\limits_{\theta \in [0,2\pi] }
g_{m,n}(\theta) $ is derived by performing a one-dimensional search and $g_{m,n}(\theta)$ is calculated by substituting the equivalent phase shift model in (\ref{fitness}) into $\lambda_3 K |\beta(\theta)|^2 -2\mathcal R\{ c_{m,n}\beta(\theta)e^{j\theta}\}$, whose expression is given by
\begin{align}\label{def:g_mn_theta}
&\ g_{m,n}(\theta) \nonumber\\
\!=&\ \lambda_3K\! \left( \xi^2(\sin(\theta\!-\!\delta)\!+\!1)^{2\alpha} \!+\! \beta_\text{min}^2 \!+\! 2\xi \beta_\text{min} (\sin(\theta\!-\!\delta)\!+\!1)^\alpha   \right) \nonumber\\
&\! -\!\! 2|c_{m,n}| \! (\xi(\sin(\theta\!-\!\delta)\!+\!1)^\alpha \!\!+\! \beta_\text{min}) \cos(\arg(c_{m,n} ) \!+\! \theta).
\end{align}

Based on the two proposed solutions, the LMMSE channel estimation algorithm is summarized in Algorithm~\ref{alg:LMMSEAO}. Regarding the convergence, it can be verified that the algorithm generates a non-increasing sequence based on the alternating optimization method that is utilized. Moreover, the objective value of the problem in (\ref{LMMSEiniProb}) has a finite lower bound. Therefore, Algorithm \ref{alg:LMMSEAO} always converges.

\begin{algorithm}[t]
\caption{Proposed MM-Based Alternating Algorithm for LMMSE Channel Estimator}
\label{alg:LMMSEAO}
\begin{algorithmic}[1]
\STATE \textit{Initialization:} Set the initial point $\mathbf{V}^{(0)}$, $\mathbf{X}^{(0)}$, iteration index $i = 0$, and convergence accuracy $\epsilon$.
\REPEAT
\STATE Set $i=i+1$.\\

    \STATE
    Obtain $\mathbf{X}^{(i)}$ according to (\ref{LMMSEXupdate}); \\

    \STATE
    Obtain $\mathbf{V}^{(i)}$ according to (\ref{LMMSEV});

\UNTIL convergence.
\STATE Output the optimal $\mathbf{V}^{(i)}$ and $\mathbf{X}^{(i)}$.
\end{algorithmic}
\end{algorithm}

\section{Accelerated MM Algorithm}
When utilizing MM algorithms, the convergence speed usually depends on the tightness of the majorization functions. In the previous sections, in order to solve the channel estimation problems and reduce the computational complexity, the coefficients of the quadratic term, e.g., $\lambda_1$, in the proposed majorization functions were relaxed and the original objective function was majorized twice successively for the LMMSE estimation. These operations result in a slow convergence speed of the MM method due to the loose surrogate function.

In this section, we employ the squared iterative method (SQUAREM) \cite{squarem} to accelerate the convergence of the proposed MM-based algorithms. SQUAREM was originally proposed in \cite{squarem} to accelerate the convergence speed of the expectation-maximization (EM) algorithms. Since MM is a generalization of EM \cite{PalomarMMAlg}, SQUAREM can also be easily applied to MM algorithms \cite{squarem1,squarem2}.

Without loss of generality, we focus on the acceleration of Algorithm \ref{alg:LSAO} for the LS channel estimation problem. The acceleration schemes of Algorithm \ref{alg:LMMSEAO} for the LMMSE channel estimator can be obtained in a similar way.
Specifically, given $\mathbf V_0$ denoting the solution obtained in the $(i-1)$-th iteration, we denote the update of $\mathbf V$ in step 5 of Algorithm \ref{alg:LSAO} by $\mathbf V^{(i)} = \text{MMupdate}(\mathbf V_0).$
Then, the proposed SQUAREM-based accelerated MM scheme is given in Algorithm \ref{alg:acceMMAlg}.
Note that the updated variable $\mathbf V_0 -2 l \mathbf L_1 + l^2 \mathbf L_2$ in step 9 may violate the constraints, i.e., $[\mathbf V_0 -2 l \mathbf L_1 + l^2 \mathbf L_2]_{m,n}\notin \mathcal F$, which needs to be projected back to the feasible region (denoted by $\mathcal P (\cdot)$).
For the element-wise practical circuit model constraint of $\mathbf V$, the projection can be readily addressed by adjusting the reflection amplitude of each element according to its phase shift based on the equivalent relationship in (\ref{fitness}). Moreover, the step length $l$ is chosen based on the Cauchy-Barzilai-Borwein (CBB) method, and a back-tracking based strategy is adopted to guarantee the monotonicity of the algorithm.

\begin{algorithm}[t]
\caption{Accelerated MM Algorithm to Optimize $\mathbf V$ for the LS Channel Estimator}
\label{alg:acceMMAlg}
\begin{algorithmic}[1]
\STATE \textit{Initialization:} set the initial point $\mathbf V^{(0)}$, iteration index $i=0$, and convergence accuracy $\epsilon$.
\REPEAT
    \STATE Set $i = i + 1$;
    \STATE $\mathbf V_1 = \text{MMupdate}\left(\mathbf V_0 \right)$;
    \STATE $\mathbf V_2 = \text{MMupdate}\left( \mathbf V_1 \right)$;
    \STATE $\mathbf L_1 = \mathbf V_1 - \mathbf V_0$;\\
    \STATE $\mathbf L_2 = \mathbf V_2 - \mathbf V_1 - \mathbf L_1$;\\
    \STATE Compute the step-length $l = - \frac{\|\mathbf L_1\|_F}{\|\mathbf L_2\|_F}$;\\
    \STATE $\mathbf V_\text{temp} = \mathcal P \left(\mathbf V_0 -2 l \mathbf L_1 + l^2 \mathbf L_2\right)$; \\
    \STATE while $\text{MSE}(\mathbf V_\text{temp} ) > \text{MSE}( \mathbf V_0)$ do \\
    $\quad \quad l \leftarrow (l - 1)/2$ \\
    $\quad \quad \mathbf V_\text{temp} = \mathcal P \left(\mathbf V_0 -2 l \mathbf L_1 + l^2 \mathbf L_2\right)$ \\
    end while\\
    \STATE $\mathbf V^{(i)} = \mathbf V_\text{temp}$;\\
\UNTIL convergence.
\STATE Output the optimal $\mathbf V^{(i)}$.
\end{algorithmic}
\end{algorithm}

For the optimization of $\mathbf X$, a similar approach as Algorithm~\ref{alg:acceMMAlg} can be applied, while only changing the projection function $\mathcal P (\cdot)$ in step 9. Specifically, the projection of $\mathbf X_0 -2 l \mathbf L_1 + l^2 \mathbf L_2$ onto the per-UE transmit power constraint set can be expressed as
$
[\mathbf X_\text{temp}]_{k,1:\tau} = \frac{\sqrt{P_k}}{\| \mathbf x'_k \|} \mathbf x'_k,\ \forall k\in \mathcal K,
$
where $\mathbf x'_k \triangleq [\mathbf X_0 -2 l \mathbf L_1 + l^2 \mathbf L_2]_{k,1:\tau}$.

\section{Simulation Results and Complexity Analysis}
\subsection{Simulation Setup}
In this section, the performance of the proposed channel estimation algorithms is evaluated via numerical simulations.
Under practical size limitations, the distance between adjacent elements at the RIS cannot be excessively large.
Hence, the channels of the RIS elements are also usually spatially correlated \cite{RIScorrelation}.
We denote the spatial correlation matrices at the UEs, RIS, and BS by
$[\mathbf \Psi]_{i,j} = \psi^{|i-j|} , \ \forall i,j,$
where $0 \leq \psi <1$ stands for the spatial correlation coefficient.
The correlation coefficients at the UEs, RIS, and BS are denoted as $\psi_\text{UE}$, $\psi_\text{RIS}$, and $\psi_\text{BS}$, respectively. The expression of the cascaded channel correlation matrix $\mathbf R_{\mathbf \Gamma}$ for the LMMSE estimator is given in Appendix \ref{channelcorrelation}.
The number of training subframes $B$ and the number of training symbols in each subframe $\tau$ are set to $M+1$ and $K$, respectively.
The noise variance $\sigma^2$ is normalized and the system SNR is defined as $\frac{P_k}{\sigma^2} = P_k$, where $P_k$ is identical for all the $K$ UEs for simplicity. The parameters in (\ref{fitness}) for modeling the practical phase shift constraints are set according to \cite{RISPracticalModel}. The simulation parameters are given in Table \ref{table:para} unless otherwise specified.
\begin{table}[t]
\caption{SIMULATION PARAMETERS}
\label{table:para}
    \centering
	\begin{tabular}{|c|c|c|}
		\hline \textbf{Notation} & \textbf{Parameter} & \textbf{Value} \\
        \hline $K$ & Number of UEs  & 4\\
        \hline $M$ & Number of RIS reflecting elements  & 20 \\
        \hline $L$ & Number of BS antennas  &  16 \\
        \hline $\psi_\text{UE}$ & Correlation coefficient at the UEs  & 0.2 \\
        \hline $\psi_\text{RIS}$ & Correlation coefficient at the RIS  & 0.4\\
        \hline $\psi_\text{BS}$ & Correlation coefficient at the BS  & 0.6 \\
        \hline $\beta_\text{min}$ & Minimum reflection amplitude in (\ref{fitness})  &  0.2 \\
        \hline $\alpha$ & Steepness of the function curve in (\ref{fitness})  &  2.0 \\
        \hline $\delta$ & Horizontal distance between $- \pi/2$ and $\beta_\text{min}$ in (\ref{fitness})  &  $0.43 \pi$ \\
        \hline $\epsilon$ & Algorithm convergence accuracy &  $10^{-3}$ \\
        \hline
	\end{tabular}%
\end{table}

\begin{figure}[t]
\begin{center}
      \epsfxsize=7.0in\includegraphics[scale=0.5]{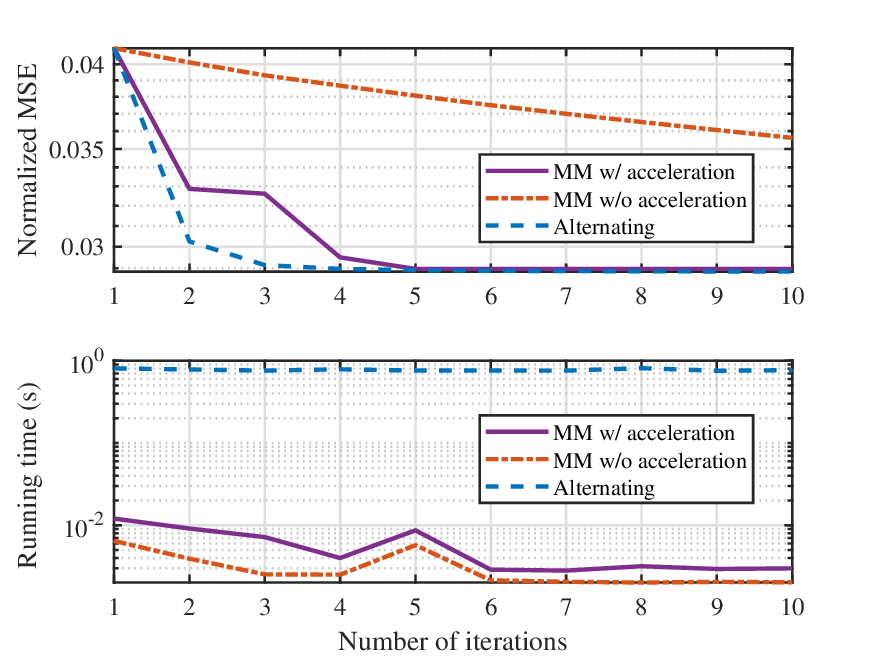}
      \caption{Normalized MSE of the LS channel estimator with respect to the number of iterations and the running time in each iteration.}\label{fig:acceMM}
    \end{center}
\end{figure}

\subsection{Normalized MSE Performance Comparisons}
We first validate the effectiveness of the utilized MM-based algorithms. We show the normalized MSE, i.e., $J_\text{LS} / [LK(M+1)]$, of the LS channel estimation algorithm with respect to the number of iterations in the first subfigure of Fig. \ref{fig:acceMM}. The alternating optimization method proposed in \cite{RISMIMO} is also used to solve problem (\ref{prob:LSV1}) for comparison, which optimizes each element of $\mathbf V$ in an alternating manner, by performing a one-dimensional search with the other $B(M+1)-1$ variables being fixed. It is found that the convergence speed of the MM algorithm is significantly enhanced by utilizing the proposed acceleration scheme, and the accelerated MM algorithm achieves almost the same MSE performance as the alternating scheme with a similar convergence speed. Moreover, as shown in the second subfigure, the MM-based algorithm has a much shorter running time than the alternating scheme.

For performance comparisons, we consider the following baseline schemes:
\subsubsection{Ideal RIS}
This corresponds to an ideal RIS, where the amplitude of the reflection coefficient of each element is fixed to $1$ while the phase shift can take any values from $0$ to $2\pi$. In this case, the solution to $\mathbf V$ in each iteration becomes
\begin{align}
\left[\mathbf{V}\right]_{m,n} = e^{-j \arg(-[\mathbf A_0]_{n,m})}\ m \in \mathcal M, \ n \in \mathcal B,
\end{align}
for the LS channel estimator, and becomes
\begin{align}\label{solution:unit1}
\left[\mathbf{V}\right]_{m,n} = e^{-j \arg(c_{m,n})}\ m \in \mathcal M, \ n \in \mathcal B,
\end{align}
for the LMMSE channel estimator.
The training matrix $\mathbf X$ is determined via the proposed solutions, i.e., an orthogonal $\mathbf X$ is used for the LS channel estimator and an alternating optimization algorithm based on (\ref{solution:unit1}) and Algorithm~\ref{alg:LMMSEAO} is utilized for the LMMSE channel estimator.

\subsubsection{Ideal RIS projection}
This corresponds to projecting the reflection coefficients obtained by the ``Ideal RIS'' strategy onto the practical constraints in (\ref{fitness}), i.e.,
$
\mathbf{\hat{V}}_{m,n} = \beta(\theta^\star_{m,n}) e^{j \theta^\star_{m,n}}, \ m \in \mathcal M, \ n \in \mathcal B,
$
where $\mathbf{\hat{V}}_{m,n}$ is the projected reflection coefficient and $\theta^\star_{m,n}$ denotes the phase shift of the ``Ideal RIS'' scheme. The training symbols are also obtained utilizing the proposed algorithms.

\subsubsection{Naive scheme}
This corresponds to computing an orthogonal training matrix according to (\ref{LSoptX}), and to deriving a reflection pattern $\mathbf{\bar{V}}$ by projecting each entry of a $B \times B$ DFT matrix onto the practical constraints in (\ref{fitness}), i.e.,
$
\mathbf{\bar{V}}_{m,n} = \beta(d_{m,n}) e^{j d_{m,n}}, \ m \in \mathcal M, \ n \in \mathcal B,
$
where $d_{m,n}$ denotes the phase shift of the $(m,n)$-th entry of the considered DFT matrix.
This naive scheme avoids the optimization process, and it is also utilized as the initial point for the proposed iterative algorithms.

\subsubsection{On-off scheme}
This corresponds to turning one RIS element with unit amplitude reflection coefficient and to estimating the associated effective channel at a time \cite{onoff}.

\begin{figure}[t]
\begin{center}
      \epsfxsize=7.0in\includegraphics[scale=0.5]{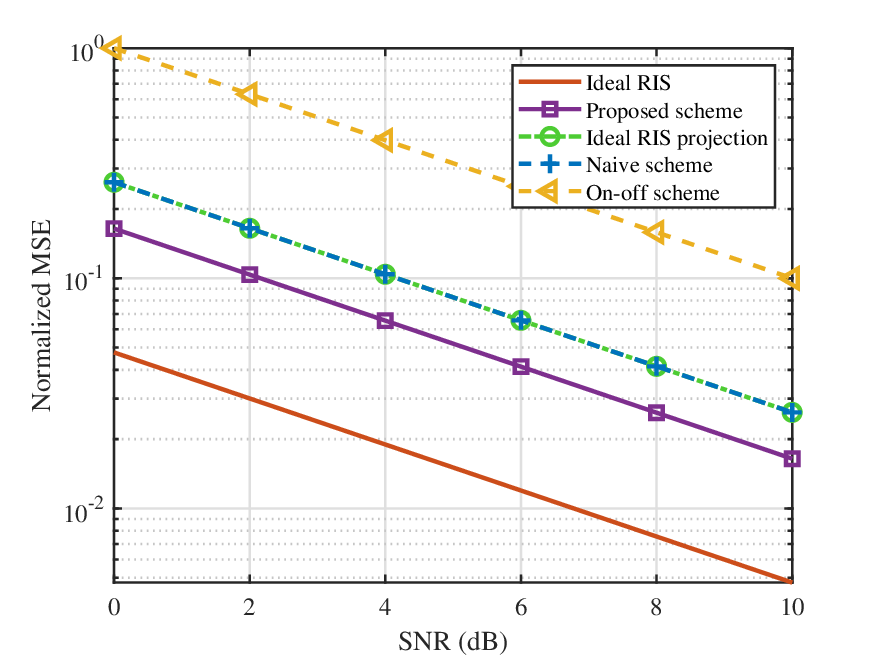}
      \caption{Normalized MSE of the LS channel estimator versus the SNR.}\label{fig:LS1}
    \end{center}
\end{figure}

\begin{figure}[t]
\begin{center}
      \epsfxsize=7.0in\includegraphics[scale=0.5]{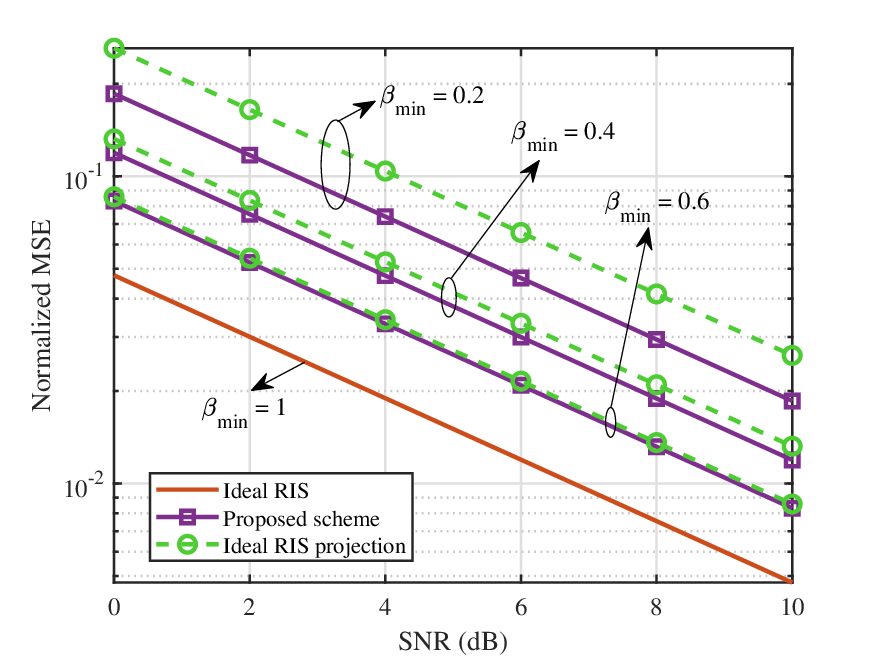}
      \caption{Normalized MSE of the LS channel estimator under different $\beta_\text{min}$.}\label{fig:LS2}
    \end{center}
\end{figure}

Fig. \ref{fig:LS1} shows the channel estimation error of the LS estimator versus the SNR. It is seen that the normalized MSEs of all the schemes decrease when increasing the SNR. Compared to the ``On-off scheme'', the other schemes, where all the RIS elements are turned on during the training phase and the reflection pattern is appropriately configured, achieve much better channel estimation performances.
Moreover, the proposed scheme outperforms the ``Ideal RIS projection'' scheme, since the proposed scheme optimizes the reflection coefficients of the RIS by incorporating the practical phase shift model directly, while the ``Ideal RIS projection'' scheme cannot guarantee any optimality when the practical phase shift model is considered.
On the other hand, we observe that the performance achieved by ``Ideal RIS projection'' and ``Naive scheme'' is similar at various SNRs. This is because a DFT-based reflection pattern is already optimal for the LS channel estimator under the ideal unit-modulus coefficient constraints \cite{fullpattern}.
Fig. \ref{fig:LS2} illustrates the impact of $\beta_\text{min}$ for the practical phase shift model.
As $\beta_\text{min}$ decreases, the phase shift model in (\ref{fitness}) deviates from the ideal one, so that the channel estimation performance degrades. Additionally, when $\beta_\text{min}$ increases, the performance gap between the ``Ideal RIS projection'' and ``Proposed scheme'' becomes smaller, since the considered phase shift model approaches the ideal one for large $\beta_\text{min}$.

\begin{figure}[t]
\begin{center}
      \epsfxsize=7.0in\includegraphics[scale=0.5]{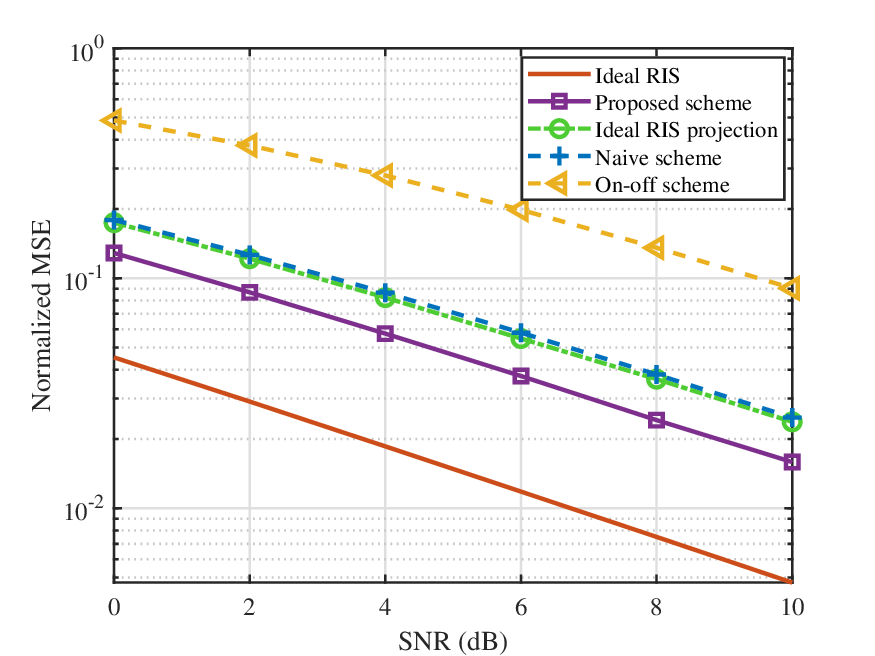}
      \caption{Normalized MSE of the LMMSE channel estimator versus the SNR.}\label{fig:LMMSE_versusSNR}
    \end{center}
\end{figure}

\begin{figure}[t]
\begin{center}
      \epsfxsize=7.0in\includegraphics[scale=0.5]{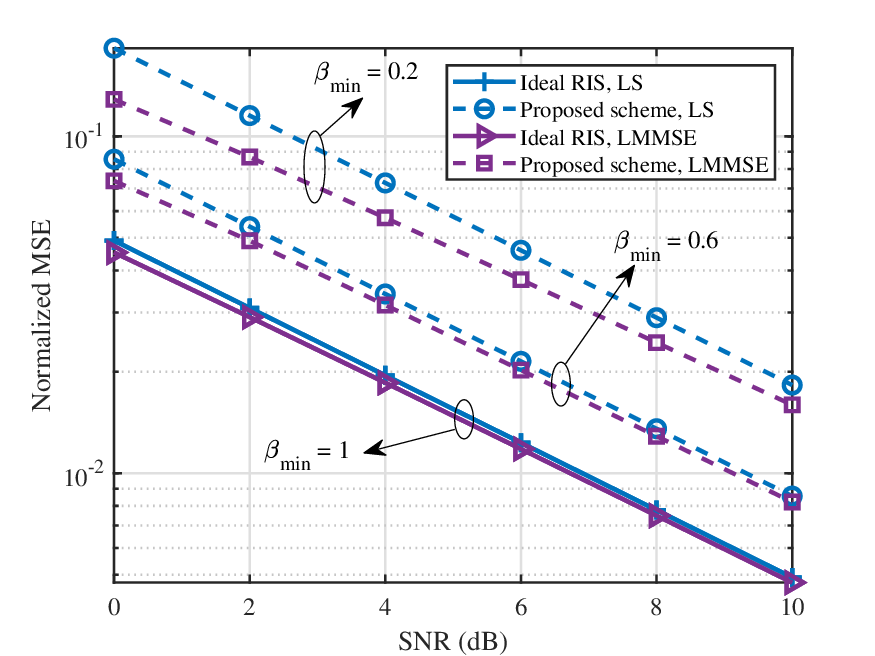}
      \caption{Normalized MSEs of the LS and LMMSE channel estimators versus the SNR.}\label{fig:comparison}
    \end{center}
\end{figure}

Fig. \ref{fig:LMMSE_versusSNR} evaluates the estimation error of the LMMSE channel estimator, from which similar conclusions can be drawn as in Fig. \ref{fig:LS1}, except that there exists a performance gap between the ``Ideal RIS projection'' and ``Naive scheme'' in Fig. \ref{fig:LMMSE_versusSNR}. This is due to the fact that, in contrast to the LS criterion, a DFT-based orthogonal reflection pattern is no longer optimal for the LMMSE channel estimator even under the ideal unit-modulus coefficient constraint.

\begin{figure}[t]
\begin{center}
      \epsfxsize=7.0in\includegraphics[scale=0.5]{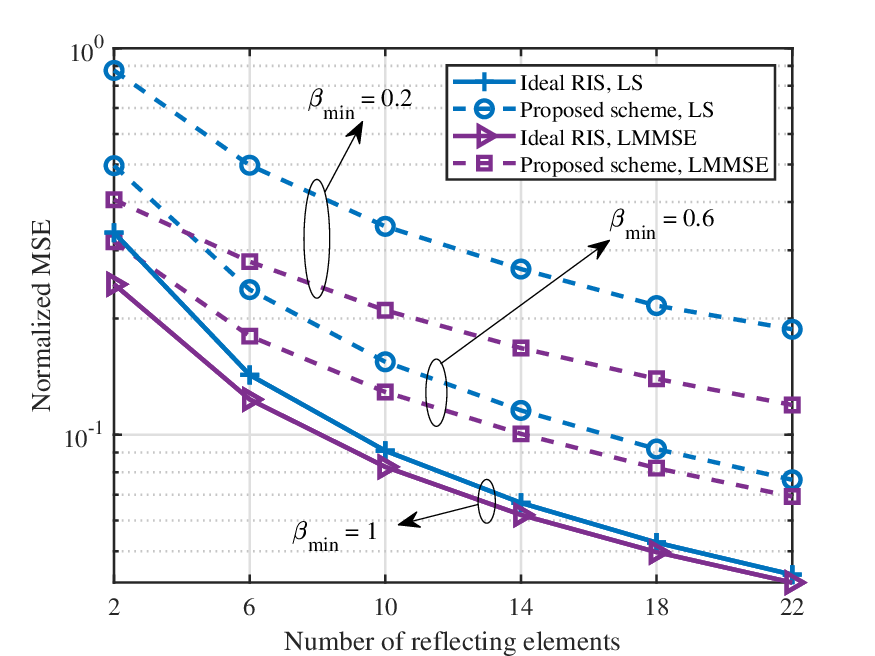}
      \caption{Normalized MSEs of the LS and LMMSE channel estimators versus the number of reflecting elements at the RIS (SNR = 0~dB).}\label{fig:comparisonM}
    \end{center}
\end{figure}

Finally, we compare the normalized MSE performance of the LS and LMMSE channel estimators for the considered RIS-aided multiuser system, as illustrated in Fig.~\ref{fig:comparison} and Fig.~\ref{fig:comparisonM}.
We observe that, by using some prior knowledge of the channel correlation, the LMMSE channel estimator is capable of achieving a lower MSE than the LS estimator.
In particular, compared to the ideal phase shift model, the performance gap between the LS and LMMSE channel estimators becomes larger when considering the non-ideal phase shift model (see Appendix~\ref{NMSEgap} for a more detailed explanation).
In addition, we find from Fig.~\ref{fig:comparisonM} that the normalized MSE of the considered system decreases when increasing the number of RIS reflecting elements. This is because the reflection pattern is well designed and the RIS can be fully exploited, which, however, requires a longer pilot sequence and leads to a higher training overhead. Finally, a more notable performance gap between the LMMSE and LS channel estimators can be observed when the number of RIS reflecting elements is relatively small.

\subsection{Impact of Channel Estimation Training Overhead}
In this subsection, we evaluate the impact of the channel estimation training overhead. Without loss of generality, we consider a block fading channel model where the instantaneous CSI remains static within each coherence block of $T$ slots and the long-term CSI, i.e., the channel correlation, remains unchanged during $U$ coherence blocks.
For comparisons, we consider two low-overhead schemes: the grouping scheme \cite{onoffgroup} and the two-timescale scheme \cite{twotimescale1,twotimescale2}.
In the former scheme, $\rho \geq 1$ neighboring RIS elements are grouped, and they share a common reflection coefficient. Accordingly, the channel estimator provides estimates for the combined channel of each group.
In this way, the effective dimension of the RIS is reduced to $M/\rho$. In the latter scheme, the reflection pattern at the RIS is designed and fixed within $U$ coherence blocks and the effective instantaneous BS-UE channel is estimated in each coherence block, with training overhead $K$, and then used for BS beamforming. We summarize and compare the training overhead and the implementation cost of these schemes in Table \ref{table1}.
It can be seen that the training overhead of the proposed method can be reduced via the RIS element grouping and the two-timescale scheme has the lowest training overhead and implementation cost.

\begin{table*}[t]
\caption{Training Overhead and Implementation Cost Comparison Among Different Schemes}
\label{table1}
    \centering
	\begin{tabular}{|c|c|c|c|}
		\hline \textbf{Scheme} & \textbf{Estimated channel} & \textbf{Training overhead} &\textbf{Implementation cost of configuring the reflection pattern} \\
        \hline Proposed w/o grouping & $[\mathbf \Gamma_1,\cdots, \mathbf \Gamma_{M+1} ]$  & $K(M+1)$ &  \textit{Training:} Calculated once every $U$ blocks; Adjusted once every $K$ slots\\
        \cline{1-3}
        {Proposed w/ grouping} & $[\mathbf {\bar\Gamma}_1,\cdots, \mathbf {\bar\Gamma}_{M/\rho+1} ]$ & $K(M/\rho+1)$ &  \textit{Transmission:} Calculated once and kept fixed within each coherence block  \\
        \hline {Two-timescale} & $\mathbf G \mathbf \Phi \mathbf H_r + \mathbf H_d $  & $K$ &  Calculated once and kept fixed within $U$ coherence blocks  \\
        \hline
	\end{tabular}%
\end{table*}

Next, we compare the average transmission rate of the schemes in Table \ref{table1}.
Note that the two-timescale design for a multiuser system in the presence of spatially correlated channels and a non-ideal phase-shift response is very challenging to analyze and, to the best of the authors' knowledge, it has not been studied in the existing literature.
Hence, we focus on a single-UE scenario for this performance comparison.
Specifically, we denote the BS-RIS channel, the RIS-UE channel, and the BS-UE channel in the $u$-th coherence block by $\mathbf{G}(u)$, $\mathbf{h}_r(u)$, and $\mathbf{h}_d(u)$, respectively.
According to \cite[Section III]{twotimescale1}, the optimization of the phase shift matrix $\mathbf{\Phi}$ based on long-term statistics is formulated by
\begin{align}\label{prob:tts}
\mathop \text{maximize}\limits_{\mathbf{\Phi}} \quad & \mathbb E \left\{ \| \mathbf{G}(u) \mathbf{\Phi} \mathbf{h}_r(u) + \mathbf{h}_d(u) \|^2 \right\} \nonumber \\
    \text{subject to} \quad &  \phi_{m} \in \mathcal{F}, \ m \in \mathcal M,
\end{align}
where the expectation is taken over the instantaneous CSI $\{\mathbf{G}(u),\mathbf{h}_r(u),\mathbf{h}_d(u)\}$. By noting $\mathbb E \left\{ \| \mathbf{G}(u) \mathbf{\Phi} \mathbf{h}_r(u) + \mathbf{h}_d(u) \|^2 \right\} = L \text{Tr}[ \mathbf{\Phi}^H \mathbf \Psi_\text{RIS} \mathbf{\Phi} \mathbf \Psi_\text{RIS} ]+ L$ and exploiting the first-order Taylor expansion:
$
\text{Tr}[ \mathbf{\Phi}^H \mathbf \Psi_\text{RIS} \mathbf{\Phi} \mathbf \Psi_\text{RIS} ]
\geq  2 \mathcal R \left\{ \text{Tr} \left[\mathbf \Psi_\text{RIS} \mathbf{\Phi}_0^H \mathbf \Psi_\text{RIS} \mathbf{\Phi}   \right] \right\}
 - \text{Tr}[ \mathbf{\Phi}_0^H \mathbf \Psi_\text{RIS} \mathbf{\Phi}_0 \mathbf \Psi_\text{RIS} ],
$
we address problem (\ref{prob:tts}) by iteratively maximizing $\mathcal R \left\{ \text{Tr} \left[\mathbf \Psi_\text{RIS} \mathbf{\Phi}_0^H \mathbf \Psi_\text{RIS} \mathbf{\Phi} \right] \right\}$ under the phase-shift constraint in (\ref{fitness}), where $\mathbf{\Phi}_0$ denotes the solution obtained in the previous iteration. In each iteration, the problem has an element-wise optimal solution
$\phi_{m} = \beta(\theta_{m}) e^{j\theta_{m}}, \ m \in \mathcal M$, where
$\theta_{m}$ is obtained by performing a one-dimensional search over $\theta \in [0,2\pi]$ to maximize $|[\mathbf \Psi_\text{RIS} \mathbf{\Phi}_0^H \mathbf \Psi_\text{RIS}]_{m,m}| (\xi(\sin(\theta-\delta)+1)^\alpha + \beta_\text{min}) \cos(\arg([\mathbf \Psi_\text{RIS} \mathbf{\Phi}_0^H \mathbf \Psi_\text{RIS}]_{m,m} ) + \theta)$.
After determining the optimal $\mathbf{\Phi}$, which is kept fixed within $U$ coherence blocks, a pilot symbol is transmitted from the UE to estimate the instantaneous effective BS-UE channel in each coherence block, and, subsequently, the maximum ratio transmission strategy is utilized at the BS for data transmission.
As a result, the average transmission rate during $U$ coherence blocks is given by $\left(1-\frac{1}{T} \right)\frac{1}{U} \sum_{u=1}^U R_\text{tts}(u)$, where $R_\text{tts}(u)$ denotes the transmission rate in the $u$-th coherence block.
As for the proposed scheme, in each coherence block a pilot sequence of length $M+1$ is transmitted from the UE to estimate the cascaded channel, by using the proposed channel estimation method, based on which the reflection coefficient matrix at the RIS and the beamforming at the BS are optimized using the method proposed in \cite{RISPracticalModel} for data transmission. Denoting the resulting transmission rate in the $u$-th coherence block by $R_\text{prop} (u)$, the average rate of the proposed scheme is given by $\left(1-\frac{M+1}{T} \right)\frac{1}{U} \sum_{u=1}^U R_\text{prop} (u)$.

\begin{figure}[t]
\begin{center}
      \epsfxsize=7.0in\includegraphics[scale=0.5]{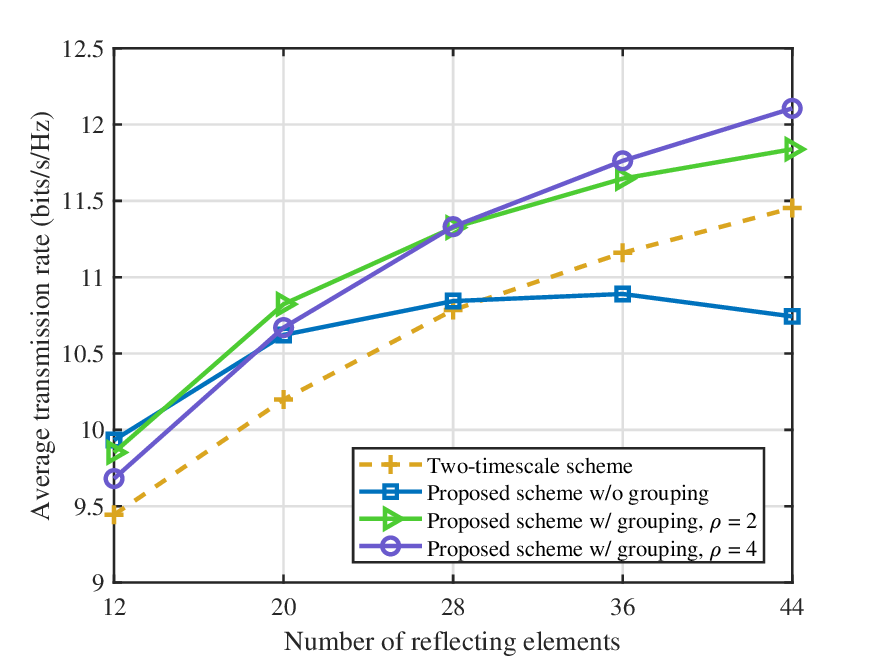}
      \caption{Average transmission rate of different schemes.}\label{fig:grouping_rate}
    \end{center}
\end{figure}

The average transmission rates of the schemes in Table \ref{table1} are compared in Fig. \ref{fig:grouping_rate}, where $U = 50$, $T = 196$, and the SNRs for channel estimation and data transmission are both set to 5~dB.
It is seen that, compared to the two-timescale scheme, the proposed scheme achieves a higher rate when $M$ is relatively small, since the RIS reflection matrix in the two-timescale scheme is predetermined and fixed within $U$ blocks while the proposed scheme adjusts the RIS reflection matrix based on the instantaneous CSI. By increasing $M$, the rate of the proposed scheme without grouping first increases and then decreases, and finally becomes worse than that of the two-timescale scheme, which is due to the larger training overhead. Nevertheless, the training overhead of the proposed scheme can be reduced by utilizing the grouping scheme. In particular, the larger $M$, the better the grouping scheme.

\subsection{Complexity Analysis}
As for the LS channel estimator, $\mathbf V$ is iteratively updated by employing Algorithm~\ref{alg:LSAO}. In each iteration, the main computational complexity lies in calculating the surrogate function $f(\mathbf V;\mathbf V_0)$ in (\ref{up:LSV}). The associated complexity is $\mathcal O (M^2B)$ and $\mathcal O (M^3)$ in terms of matrix multiplications and matrix inversions, respectively, and $\mathcal O (MB \mathcal D)$ in terms of performing the element-wise one-dimensional search for $MB$ elements according to (\ref{LSThetaUpdateRule}), where $\mathcal D$ denotes the complexity of the one-dimensional search. Hence, the overall complexity of the LS channel estimator is $\mathcal O (\mathcal I_\text{LS}(M^3 + M^2B+ MB \mathcal D))$, where $\mathcal I_\text{LS}$ denotes the number of iterations required for convergence.
As for the LMMSE channel estimator, the matrices $\mathbf X$ and $\mathbf V$ need to be updated in each iteration of Algorithm~\ref{alg:LMMSEAO}. The process of updating $\mathbf X$ involves the calculation of matrix multiplications, with complexity $\mathcal O (B^2 \tau^2MK + M^2K^2 B\tau)$, and matrix inversions, with complexity $\mathcal O (B^3 \tau^3)$, as well as the largest eigenvalue of a $BK \times BK$ matrix, which can be handled via the efficient power iteration method \cite{power_ite} with complexity $\mathcal O (B^2 K^2)$. Hence, the total complexity of updating $\mathbf X$ is $\mathcal O (B^3 \tau^3 + B^2 K^2 + B^2 \tau^2MK + M^2K^2 B\tau)$.
The update of $\mathbf V$ has an additional computational cost of $\mathcal O (MB \mathcal D)$ because of the element-wise phase search. Hence, the overall complexity of the LMMSE channel estimator is given by $\mathcal O (\mathcal I_\text{LMMSE}(B^3 \tau^3 + B^2 K^2 + B^2 \tau^2MK + M^2K^2 B\tau + MB \mathcal D))$, where $\mathcal I_\text{LMMSE}$ denotes the number of iterations required for convergence.
Considering a typical setup, where $B = M+1$ and $\tau = K$, the total complexities of the LS and the LMMSE channel estimation schemes are $\mathcal O (\mathcal I_\text{LS}(M^3 + M^2 \mathcal D))$ and $\mathcal O (\mathcal I_\text{LMMSE}(M^3 K^3 + M^2\mathcal D))$, respectively. Hence, the LMMSE channel estimator has a higher computational complexity than the LS channel estimator.

\section{Conclusion}
A joint design for training symbols and reflection pattern for RIS-assisted multiuser communication systems with a realistic phase-amplitude reflection model was investigated in this paper. We considered the MSE minimization problem for both LS and MMSE channel estimators, subject to the transmit power constraint at the UEs and a practical phase shift model at the RIS.
For the LS criterion, we proved the optimality of the orthogonal training signals and developed an MM-based algorithm to address the reflection pattern design with a semi-closed form solution in each iteration. As for the LMMSE criterion, we proposed to iteratively optimize the training symbols and the reflection pattern, whose optimal solutions in each iteration were obtained in a closed form and a semi-closed form, respectively. The SQUAREM method was further utilized to accelerate the convergence speed of the proposed MM algorithms. Simulation results confirmed that the proposed design can effectively improve the channel estimation performance of RIS-aided channel in the presence of practical phase-amplitude reflection models.
In addition, compared to the two-timescale design scheme, the proposed instantaneous channel estimation-based scheme demonstrates superior transmission rates for a low-to-medium size of the RIS, while a grouping strategy is needed when considering a large-size RIS.

\begin{appendices}
\section{Proof of Theorem ~\ref{theorem1}}\label{proof:Theorem1}
Based on the expression of $\mathbf S$ in (\ref{def:widetilde}), we have
\begin{align}
\mathbf S \mathbf S^H =&\ [(\mathbf{V} \otimes \mathbf I_K)  (\mathbf I_B \otimes \mathbf{X})][ (\mathbf I_B \otimes \mathbf{X}^H) (\mathbf{V}^H \otimes \mathbf I_K)] \nonumber\\
=&\ (\mathbf{V} \otimes \mathbf{X}) (\mathbf{V}^H \otimes \mathbf{X}^H)  \nonumber\\
=&\ \left(\mathbf V \mathbf V^H\right) \otimes \left(\mathbf X \mathbf X^H\right). \label{VotimesX}
\end{align}
Substituting (\ref{VotimesX}) into $J_\text{LS}$, we have
\begin{align}\label{traceSS}
\text{Tr}\left[\left( \mathbf S \mathbf S^H \right)^{-1}\right] = &\
\text{Tr}\left[\left(  \left(\mathbf V \mathbf V^H\right) \otimes \left(\mathbf X \mathbf X^H\right)  \right)^{-1}\right] \nonumber \\
=&\ \text{Tr}\left[  \left(\mathbf V \mathbf V^H\right)^{-1} \otimes \left(\mathbf X \mathbf X^H\right)^{-1}  \right] \nonumber \\
=&\ \text{Tr}\left[  \left(\mathbf V \mathbf V^H\right)^{-1} \right] \text{Tr}\left[  \left(\mathbf X \mathbf X^H\right)^{-1}  \right].
\end{align}
Therefore, given an arbitrary reflection pattern $\mathbf V$, the optimization of $\mathbf{X}$ amounts to
\begin{align}\label{prob:LSX}
\mathop \text{minimize}\limits_{\mathbf X} \quad & \kappa \text{Tr}\left[  \left(\mathbf X \mathbf X^H\right)^{-1}\right] \nonumber \\
\text{subject to} \quad & \|\mathbf x_k\|^2 \leq P_k,\ k \in \mathcal K,
\end{align}
where $\kappa = \sigma^2 L  \text{Tr}[(\mathbf V \mathbf V^H)^{-1}]$ is independent of $\mathbf X$.
It can be readily shown by contradiction that the inequality constraints in (\ref{prob:LSX}) must be active at the optimality. Moreover, the optimal solution must have orthogonal rows. Hence, we obtain (\ref{optimalX}) and the proof is completed.

\section{Proof of Proposition~\ref{LSlemma1}}\label{proof1}
Denote the objective function of problem (\ref{prob:LSV1}) by $f(\mathbf{V}) \triangleq \text{Tr}[( \mathbf{V} \mathbf{V}^H)^{-1}]$. To find a proper surrogate function of $f(\mathbf{V})$, we utilize the following upper bound \cite[Eq. (25)]{PalomarMMAlg}
\begin{align}\label{MMbound}
f(\mathbf x) \leq f(\mathbf y) + \nabla f(\mathbf y)^T(\mathbf x- \mathbf y) + \frac{1}{2}(\mathbf x- \mathbf y)^T \mathbf M (\mathbf x- \mathbf y),
\end{align}
where the matrix $\mathbf M$ must satisfy $\mathbf M \succeq \bigtriangledown^2 f(\mathbf x)$ for all $\mathbf x$.
According to (\ref{MMbound}), we are ready to calculate the first-order and the second-order differentials of $f(\mathbf{V})$. By applying $d \mathbf A^{-1} = - \mathbf A^{-1} (d \mathbf A) \mathbf A^{-1}$ \cite{Complex-Valued}, we first compute the first-order differential of $f(\mathbf{V})$ as
\begin{align}\label{firstorder}
 d f(\mathbf{  V})
=& -\text{Tr}\left[\left( \mathbf{  V} \mathbf{  V}^H \right)^{-2} d\left( \mathbf{  V} \mathbf{  V}^H \right) \right]  \nonumber \\
= & -\text{Tr}\left[ \mathbf{  V}^H  \left(\mathbf{  V} \mathbf{  V}^H\right)^{-2}  d\mathbf{ {V}}
+  \left(\mathbf{  V}  \mathbf{  V}^H\right)^{-2}  \mathbf{  V}  d \mathbf{ {V}}^H
\right].
\end{align}
Then, according to \cite{Complex-Valued} and $\text{Tr}(\mathbf A \mathbf B)= \text{vec}^T(\mathbf A^T) \text{vec}(\mathbf B)$, we obtain the second-order differential of $f(\mathbf{V})$ by
\begin{align}\label{seconddifferential}
d^2 f(\mathbf{  V})
=&\ -d \text{vec}^T\left( \left(\mathbf{  V} \mathbf{  V}^H\right)^{-2T} \mathbf{ {V}}^* \right)   d \text{vec}\left( \mathbf{ {V}} \right) \nonumber\\
&\ - d \text{vec}^T\left( \left(\mathbf{  V} \mathbf{  V}^H\right)^{-2}   \mathbf{  V} \right)  d \text{vec}\left( \mathbf{ {V}}^*  \right).
\end{align}
Subsequently, we manipulate the first term of (\ref{seconddifferential}) as follows:
\begin{align}\label{LS1d1}
&\ -d \text{vec}^T\left( \left(\mathbf{  V} \mathbf{  V}^H\right)^{-2T} \mathbf{ {V}}^* \right)   d \text{vec}\left( \mathbf{ {V}} \right) \nonumber \\
=&\  - \left[\text{vec}^T\left( d \left(\mathbf{  V} \mathbf{  V}^H\right)^{-T}   \left(\mathbf{  V} \mathbf{  V}^H\right)^{-T} \mathbf{ {V}}^* \right) \nonumber \right.\\
&\ \quad \left. +\text{vec}^T\left( \left(\mathbf{  V} \mathbf{  V}^H\right)^{-T}   d \left(\mathbf{  V} \mathbf{  V}^H\right)^{-T} \mathbf{ {V}}^*\right)
\nonumber \right.\\
&\ \quad \left. +\text{vec}^T\left( \left(\mathbf{  V} \mathbf{  V}^H\right)^{-2T} d \mathbf{ {V}}^* \right)\right]
d \text{vec}\left(\mathbf{ {V}}\right) \nonumber\\
=&\ \left[\text{vec}^T\left( \left(\mathbf{  V} \mathbf{  V}^H\right)^{-T}
(\mathbf{  V}^*  d \mathbf{  V}^T \!\!+\!  d \mathbf{  V}^*  \mathbf{  V}^T)\!
\left(\mathbf{  V} \mathbf{  V}^H\right)^{-2T}  \mathbf{ {V}}^* \right)
\nonumber \right.\\
&\ \left. +\text{vec}^T\!\left( \left(\mathbf{  V}\mathbf{  V}^H\right)^{-2T}\!
(\mathbf{  V}^*  d \mathbf{  V}^T \!\!+\!  d \mathbf{  V}^* \mathbf{  V}^T)\!
 \left(\mathbf{  V}  \mathbf{  V}^H\right)^{-T} \mathbf{ {V}}^*  \right)
\nonumber \right.\\
&\ \left. - \text{vec}^T\left( \left(\mathbf{  V} \mathbf{  V}^H\right)^{-2T}  d \mathbf{ {V}}^* \right)\right]
d \text{vec}\left(\mathbf{ {V}}\right) \nonumber\\
\triangleq &\ d \text{vec}^T\!\left(\mathbf { {V}}^*\right) \mathbf{P} d \text{vec}\left(\mathbf{ {V}}\right)
\!+\! d \text{vec}^T\!\left(\mathbf { {V}}\right) \mathbf K^T \mathbf{Q}  d \text{vec}\left(\mathbf{ {V}}\right),
\end{align}
where the last equality holds based on the relationship $ \text{vec}\left( \mathbf A \mathbf B \mathbf C \right) = (\mathbf C^T \otimes \mathbf A) \text{vec}\left( \mathbf B \right)$, $\mathbf K$ is the unique $[(M+1)B] \times [(M+1)B]$ permutation matrix satisfying $\mathbf K \text{vec}\left(\mathbf{A}\right) = \text{vec}\left(\mathbf{A}^T\right)$ for an arbitrary matrix $\mathbf{A} \in \mathbb{C}^{(M+1) \times B}$, and $\mathbf{P}$ and $\mathbf{Q}$ are defined as follows:
\begin{align}\label{def:PQAppendix}
 \mathbf{P} \triangleq&\  \mathbf{  V}^T \left(\mathbf{  V} \mathbf{  V}^H\right)^{-2T}  \mathbf{  V}^*  \otimes \left(\mathbf{  V} \mathbf{  V}^H\right)^{-1} \nonumber\\
&\  + \mathbf{  V}^T \left(\mathbf{  V} \mathbf{  V}^H\right)^{-T} \mathbf{  V}^* \otimes \left(\mathbf{  V}   \mathbf{  V}^H\right)^{-2}
 - \mathbf I_{B} \otimes \left(\mathbf{  V} \mathbf{  V}^H\right)^{-2} \nonumber\\
\mathbf{Q} \triangleq&\ \left(\mathbf{  V} \mathbf{  V}^H\right)^{-2T}  \mathbf{ {V}}^*  \otimes  \mathbf{  V}^H \left(\mathbf{  V} \mathbf{  V}^H\right)^{-1} \nonumber\\
&\ + \left(\mathbf{  V}  \mathbf{  V}^H\right)^{-T} \mathbf{ {V}}^* \otimes \mathbf{  V}^H \left(\mathbf{  V} \mathbf{  V}^H\right)^{-2}.
\end{align}
Similar to (\ref{LS1d1}), the second term of (\ref{seconddifferential}) can be written as
\begin{align}\label{LS1d2}
\!\!\! d \text{vec}^T\!\left(\mathbf{ {V}} \right)  \mathbf{P}^T d \text{vec}\!\left( \mathbf{ {V}}^*\right)
\!+\! d \text{vec}^T\!\left(\mathbf{ {V}}^* \right)   \mathbf{Q}^H \mathbf K d \text{vec}\!\left( \mathbf{ {V}}^* \right).
\end{align}
Combining the results in (\ref{LS1d1}) and (\ref{LS1d2}), we reexpress the second-order differential of $f(\mathbf{V}) $ as
\begin{align}\label{secondorder}
\!\!\!\!d^2 f(\mathbf{  V})
=& \begin{bmatrix} d \text{vec}^T\left(\mathbf{ {V}}^*\right) & d \text{vec}^T\left( \mathbf{ {V}}  \right) \end{bmatrix}
\mathcal H_{\mathbf V}
\begin{bmatrix} d \text{vec}\left(\mathbf{ {V}}  \right) \\  d \text{vec}\left(  \mathbf{ {V}}^*  \right) \end{bmatrix},
\end{align}
where
$
\mathcal H_{\mathbf V} \triangleq \begin{bmatrix} \mathbf{P} & \mathbf{Q}^H \mathbf K \\ \mathbf{K}^T\mathbf{Q} & \mathbf{P}^T \end{bmatrix}.
$
The remaining step is to find a matrix $\mathbf M$ such that $\mathbf M \succeq \mathcal H_{\mathbf V}$ holds for all feasible $\mathbf{V}$. For convenience, we simply choose $\mathbf M = \eta \mathbf I_{2(M+1)B}$ with $\eta \geq \lambda_\text{max} (\mathcal H_{\mathbf V})$.
To determine $\eta$, we first obtain the following upper bound to $\lambda_\text{max} (\mathcal H_{\mathbf V})$:
\begin{align}
\lambda_\text{max} (\mathcal H_{\mathbf V}) \overset{(\rm a)}\leq&\ \lambda_\text{max} \left( \begin{bmatrix} \mathbf{P} & \mathbf 0 \\ \mathbf 0 & \mathbf{P}^T \end{bmatrix} \right) + \lambda_\text{max} \left( \begin{bmatrix} \mathbf 0 & \mathbf{Q}^H \mathbf{K} \\ \mathbf{K}^T\mathbf{Q} & \mathbf 0 \end{bmatrix} \right) \nonumber\\
 \overset{(\rm b)} =&\ \lambda_\text{max}(\mathbf{P}) + \lambda_\text{max}^{0.5}( \mathbf{Q} \mathbf Q^H),
\end{align}
where the inequality (a) follows from $\lambda_\text{max} (\mathbf A + \mathbf B) \leq \lambda_\text{max} (\mathbf A ) + \lambda_\text{max} (\mathbf B)$ with $\mathbf A$ and $\mathbf B$ being Hermitian matrices and the equality (b) follows from $\lambda_\text{max} ([\mathbf 0 \ \mathbf A^H; \mathbf A \ \mathbf 0]) = \lambda_\text{max}^{0.5}(\mathbf A\mathbf A^H)$ \cite{MatrixAnalysis}.
Since the three terms in $\mathbf P$ are Hermitian matrices, we can further upper bound $\lambda_\text{max}(\mathbf{P})$ as
\begin{align}
\lambda_\text{max}(\mathbf{P}) \leq&\ \lambda_\text{max}\left( \mathbf{  V}^T \left(\mathbf{  V} \mathbf{  V}^H\right)^{-2T}  \mathbf{  V}^*  \otimes \left(\mathbf{  V} \mathbf{  V}^H\right)^{-1} \right) \nonumber\\
& + \!\lambda_\text{max}\left( \left( \mathbf{V}^T \left(\mathbf{  V} \mathbf{  V}^H\right)^{-T} \mathbf{V}^* \!-\! \mathbf I_B\right) \otimes \left(\mathbf{  V}   \mathbf{  V}^H\right)^{-2} \right)  \nonumber \\
=&\ \lambda_\text{max}^2\left( \left(\mathbf{V} \mathbf{V}^H\right)^{-1}\right),
\end{align}
where the equality follows from $\lambda_\text{max} (\mathbf A \otimes \mathbf B) = \lambda_\text{max} (\mathbf A)\lambda_\text{max} (\mathbf B)$, $\lambda_\text{max} (\mathbf A \mathbf B) = \lambda_\text{max} (\mathbf B \mathbf A)$, and $\lambda_\text{max} (\mathbf A) = \lambda_\text{max} (\mathbf A^T)$.
With similar procedures, $\lambda_\text{max}^{0.5}( \mathbf{Q}\mathbf{Q}^H)$ can be upper bounded by $ 2 \lambda_\text{max}^2( (\mathbf{V} \mathbf{V}^H)^{-1})$. Thus, we conclude that $\lambda_\text{max} (\mathcal H_{\mathbf V}) \leq 3 \lambda_\text{max}^2((\mathbf{V} \mathbf{V}^H)^{-1})$.

However, it is still hard to determine the largest eigenvalue of $(\mathbf{V} \mathbf{V}^H)^{-1}$ for every feasible solution $\mathbf{V}$, since there exist cases where $\mathbf{V} \mathbf{V}^H$ tends to be a singular matrix and thus the largest eigenvalue of $(\mathbf{V} \mathbf{V}^H)^{-1}$ tends to be infinite.
Fortunately, we can handle this difficulty by fully exploiting the specific form of the considered problem. Specifically, we impose an additional constraint $\text{Tr}[( \mathbf{V}  \mathbf{ V}^H )^{-1}] \leq \text{Tr}[( \mathbf{\hat V}  \mathbf{\hat V}^H )^{-1}]$ to problem (\ref{prob:LSV1}) with $\mathbf{\hat V}$ denoting an arbitrary feasible solution, which yields
\begin{align}\label{prob:LSV3}
\mathop \text{minimize}\limits_{ \mathbf{V}} \quad &  \text{Tr}\left[\left( \mathbf{V}  \mathbf{ V}^H \right)^{-1}\right] \nonumber \\
\text{subject to} \quad
& [\mathbf V]_{m,n} \in \mathcal{F}, \ m \in \mathcal M, \ n \in \mathcal B \nonumber\\
& [\mathbf V]_{M+1,n} = 1, \ n \in \mathcal B \nonumber\\
& \text{Tr}\left[\left( \mathbf{V}  \mathbf{ V}^H \right)^{-1}\right] \leq \text{Tr}\left[\left( \mathbf{\hat V}  \mathbf{\hat V}^H \right)^{-1}\right].
\end{align}
It can be readily shown that problems (\ref{prob:LSV1}) and (\ref{prob:LSV3}) have the same global optimal solution. Based on the additional imposed constraint, we can relax the largest eigenvalue of $(\mathbf{V} \mathbf{V}^H)^{-1}$ as $\text{Tr}[( \mathbf{\hat V} \mathbf{\hat V}^H )^{-1}]$ and accordingly set $\eta$ to $3 \text{Tr}[( \mathbf{\hat V} \mathbf{\hat V}^H )^{-1}]^2$. Moreover, we judiciously update $\eta = 3 \text{Tr}[( \mathbf{V}_0 \mathbf{V}_0^H )^{-1}]^2$ in each iteration with $\mathbf{V}_0$ being the obtained solution in the previous iteration for facilitating a tighter bound. Note that the above operations will not affect the convergence of the MM algorithm \cite{squarem2}.

By substituting the first-order differential in (\ref{firstorder}) and $\mathbf M = 3 \text{Tr}[( \mathbf{V}_0 \mathbf{V}_0^H )^{-1}]^2 \mathbf I$ into (\ref{MMbound}), we finally obtain the surrogate function of $f(\mathbf V)$ by (\ref{up:LSV}) in Lemma \ref{LSlemma1}.

\section{Proof of Proposition ~\ref{prop:LSexhaustive}}\label{proof:LSexhaustive}
Note that the objective function of problem (\ref{prob:LSV2}) can be rewritten as $ \sum_{m=1}^{M+1} \sum_{n=1}^B f_{m,n}([\mathbf{V}]_{m,n})$, where $f_{m,n}([\mathbf{V}]_{m,n}) \triangleq \lambda_1|[\mathbf{V}]_{m,n}|^2 + 2\mathcal R \{[\mathbf{A}_0 ]_{n,m} [\mathbf{V}]_{m,n}\}$. For a fixed $(m,n)$, the value of $f_{m,n}([\mathbf{V}]_{m,n})$ depends on $[\mathbf{V}]_{m,n}$ and is independent of the other elements in $\mathbf{V}$. Together with the fact that the constraints in problem (\ref{prob:LSV2}) are imposed on each element of $\mathbf{V}$ independently, we conclude that problem (\ref{prob:LSV2}) can be solved for each element of $\mathbf{V}$ independently, i.e., in an element-wise manner.
Without loss of generality, replacing $[ \mathbf{V} ]_{m,n}$ with $\phi = \beta(\theta) e^{j\theta}$ and plugging the equivalent phase shift model in (\ref{fitness}), $f_{m,n}([\mathbf{V}]_{m,n})$ is equal to
\begin{align}\label{def:f(theta)}
&\ \lambda_1  |\phi|^2 + 2\mathcal R\left\{ [\mathbf{A}_0]_{n,m}  \phi \right\} \nonumber\\
= &\ \lambda_1 \beta(\theta)^2 + 2\mathcal R\left\{ |[\mathbf{A}_0]_{n,m}| \beta(\theta) e^{j(\arg([\mathbf{A}_0]_{n,m} ) + \theta) }\right\} \nonumber\\
=&\ \lambda_1 \!\left[ \xi^2(\sin(\theta\!-\!\delta)\!+\!1)^{2\alpha} \!+\! \beta_\text{min}^2 \!+ \! 2\xi \beta_\text{min} (\sin(\theta\!-\!\delta)\!+\!1)^\alpha   \right] \nonumber\\
& +\!\! 2|[\mathbf{A}_0]_{n,m}|\! (\xi(\sin(\theta\!-\!\delta)\!+\!1)^\alpha \!\!+\! \beta_\text{min})\! \cos(\arg([\mathbf{A}_0]_{n,m} ) \!+\! \theta) \nonumber\\
= &\ \bar f_{m,n}(\theta),
\end{align}
where the second equality follows from Euler's formula.
This implies that the minimization of $f_{m,n}([\mathbf{V}]_{m,n})$ can be addressed by performing a one-dimensional search over the phase shift $\theta \in [0, 2\pi]$.

\section{Proof of Theorem ~\ref{theorem2}}\label{proof:theorem2}
Denote the constraint set of problem (\ref{prob:LSV1}) by $\mathcal S$. We prove the convergence of Algorithm \ref{alg:LSAO} by verifying the following four conditions
according to \cite[Sec. III]{convergenceofMM}:
\begin{enumerate}
  \item $f\left(\mathbf V\right) \leq f\left(\mathbf V;\mathbf V_0\right), \ \forall  \mathbf V \in \mathcal S$;
  \item $f\left( \mathbf V_0 \right) = f\left(\mathbf V_0; \mathbf V_0 \right)$;
  \item $\nabla f\left( \mathbf V_0 \right) = \nabla f\left( \mathbf V_0; \mathbf V_0 \right)$;
  \item $f\left(\mathbf V; \mathbf V_0\right)$ is continuous in both $\mathbf V$ and $\mathbf V_0$.
\end{enumerate}
In particular, the first two conditions guarantee the convergence of the proposed MM algorithm while conditions 3) and 4) guarantee that the algorithm converges to a stationary point.

Clearly, the considered $f\left(\mathbf V; \mathbf V_0\right)$ in (\ref{up:LSV}) is a continuous function and thus condition 4) holds. Next, the utilized bound $f\left(\mathbf V; \mathbf V_0\right)$ is obtained according to (\ref{MMbound}) as shown in Appendix \ref{proof1}, where the original function $f(\mathbf x)$ is upper bounded by its second-order Taylor expansion, which is in the right-hand side of (\ref{MMbound}) and denoted as $f(\mathbf x;\mathbf y)$. It is readily verified from (\ref{MMbound}) that $f(\mathbf y) = f(\mathbf y;\mathbf y)$ and $\nabla f(\mathbf y) = \nabla f(\mathbf y;\mathbf y)$ and thus we conclude that the conditions 1), 2), and 3) hold for $f\left(\mathbf V\right)$ and $f\left(\mathbf V;\mathbf V_0\right)$. Therefore, the four conditions hold and the proof is completed.

\section{Proof of Lemma ~\ref{lemma:LMMSElemma1}}\label{proof:concave}
Let us introduce $\mathbf U \triangleq \mathbf S^H \mathbf R_{\mathbf \Gamma} \mathbf S + \sigma^2 L \mathbf I_{\tau B}$ and rewrite the objective
function $g(\mathbf S)$ in problem (\ref{LMMSEiniProb}) as $g(\mathbf U,\mathbf S) \triangleq - \text{Tr}\left[  \mathbf R_{\mathbf \Gamma} \mathbf S \mathbf U^{-1} \mathbf S^H \mathbf R_{\mathbf \Gamma}\right]$. Then, based on the fact that the matrix function $h(\mathbf A, \mathbf B) = \text{Tr}\left[ \mathbf B \mathbf A ^{-1}\mathbf B^H\right]$ is jointly convex in $\{\mathbf A, \mathbf B\}$ \cite{ConvexOptimization}, we can lower bound $h(\mathbf A, \mathbf B)$ by its first-order Taylor expansion as
\begin{align}\label{first-orderTaylor}
&\ \text{Tr}\left[ \mathbf B \mathbf A^{-1}\mathbf B^H\right] \nonumber\\
\geq&\ 2\mathcal R\left\{ \text{Tr}\left[  \mathbf A_0^{-1} \mathbf B_0^H \mathbf B \right] \right\}
  -  \text{Tr}\left[\mathbf A_0^{-1} \mathbf B_0^H \mathbf B_0 \mathbf A_0^{-1}  \mathbf A \right],
\end{align}
where $\mathbf A_0$ and $\mathbf B_0$ denote arbitrary feasible points. By substituting $\{\mathbf A, \mathbf B\}$ with $\{\mathbf U, \mathbf R_{\mathbf \Gamma} \mathbf S \}$, we obtain an upper bound for $g(\mathbf S)$ by
\begin{align} \label{first-orderTaylor}
& - \text{Tr}\left[  \mathbf R_{\mathbf \Gamma} \mathbf S \left( \mathbf S^H \mathbf R_{\mathbf \Gamma} \mathbf S + \sigma^2 L \mathbf I_{\tau B}\right)^{-1} \mathbf S^H \mathbf R_{\mathbf \Gamma}\right] \nonumber \\
\leq & - 2\mathcal R\left\{ \text{Tr}\left[  (\mathbf S_0^H \mathbf R_{\mathbf \Gamma} \mathbf S_0 + \sigma^2 L \mathbf I_{\tau B})^{-1} \mathbf S_0^H \mathbf R_{\mathbf \Gamma} \mathbf R_{\mathbf \Gamma} \mathbf S \right] \right\} \nonumber \\
& +  \text{Tr}\left[(\mathbf S_0^H \mathbf R_{\mathbf \Gamma} \mathbf S_0 + \sigma^2 L \mathbf I_{\tau B})^{-1} \mathbf S_0^H \mathbf R_{\mathbf \Gamma}  \mathbf R_{\mathbf \Gamma} \mathbf S_0   \right.  \nonumber \\
&\quad\quad\ \left.  \times (\mathbf S_0^H \mathbf R_{\mathbf \Gamma} \mathbf S_0 + \sigma^2 L \mathbf I_{\tau B})^{-1}(\mathbf S^H \mathbf R_{\mathbf \Gamma} \mathbf S + \sigma^2 L \mathbf I_{\tau B}) \right],
\end{align}
which is equal to $g(\mathbf S;\mathbf S_0)$ given in Lemma \ref{lemma:LMMSElemma1}.

\section{Proof of Proposition ~\ref{prop:LMMSEX}}\label{proof:KKT1}
The Karush-Kuhn-Tucker (KKT) conditions of problem (\ref{prob:LMMSperUE}) are given as follows:
\begin{align}
\| \mathbf x_k^\star \|^2 - P_k \leq &\ 0,   \label{KKTa} \\
\mu^\star \geq &\ 0,  \label{KKTb} \\
\mu^\star (\| \mathbf x_k^\star \|^2 - P_k) = &\ 0, \label{KKTc} \\
2 \lambda_2 B\mathbf x_k^\star - 2\mathbf{b}_k + 2 \mu^\star \mathbf{x}_k^\star =&\ \mathbf{0}, \label{KKTd}
\end{align}
where $\mathbf x_k^\star$ is the optimal solution to $\mathbf x_k$ and $\mu^\star$ is the optimal dual variable associated with the constraint $\|\mathbf x_k\|^2 \leq P_k $. We now analyze the KKT conditions to find $\mathbf x_k^\star$ and $\mu^\star$.

\subsubsection{Case 1}
If $\mu^\star > 0$, we obtain $\| \mathbf x_k^\star \|^2 = P_k$ from (\ref{KKTc}). Substituting this into (\ref{KKTd}), we have $\| \mathbf x_k^\star \|^2 = \frac{\| \mathbf b_k \|^2}{(\lambda_2 B + \mu^\star)^2} = P_k$, which yields $\mu^\star + \lambda_2 B= \frac{\| \mathbf b_k \|}{\sqrt{P_k}} $. Then, the optimal $\mathbf x_k^\star$ is
\begin{align}
\mathbf x_k^\star = \frac{\mathbf b_k}{\lambda_2 B + \mu^\star} =\frac{\sqrt{P_k}}{\| \mathbf b_k \|} \mathbf b_k.
\end{align}
To guarantee a positive $\mu^\star$, we must have $\| \mathbf b_k \| > \sqrt{P_k}\lambda_2 B$.

\subsubsection{Case 2}
If $\mu^\star = 0$, it follows from (\ref{KKTd}) that
\begin{align}
\mathbf{x}_k^\star = \frac{\mathbf b_k}{\lambda_2 B}.
\end{align}
By substituting this equality into (\ref{KKTa}), we have $\| \mathbf x_k^\star \|^2 =  \frac{\| \mathbf b_k \|^2}{ (\lambda_2 B)^2} \leq  P_k$.
Combining these two cases, we derive the optimal solution in (\ref{LMMSEXupdate}).

\section{Expression of the Cascaded Channel Correlation}\label{channelcorrelation}
For the purpose of modeling the cascaded channel correlation matrix $\mathbf R_{\mathbf \Gamma}$, we consider the Kronecker channel model \cite{ChannelPsi}:
$\mathbf{H}_r = \mathbf \Psi^\frac{1}{2}_{\text{RIS},\mathbf{H}_r} \mathbf{\bar H}_r  \mathbf \Psi^\frac{T}{2}_{\text{UE},\mathbf{H}_r},$
$\mathbf{G} = \mathbf \Psi^\frac{1}{2}_{\text{BS},\mathbf{G}} \mathbf{\bar G}  \mathbf \Psi^\frac{T}{2}_{\text{RIS},\mathbf{G}},$ and $\mathbf{H}_d = \mathbf \Psi^\frac{1}{2}_{\text{BS},\mathbf{H}_d} \mathbf{\bar H}_d  \mathbf \Psi^\frac{T}{2}_{\text{UE},\mathbf{H}_d}.$ The elements of the matrices $\mathbf{\bar H}_r$, $\mathbf{\bar G}$, and $\mathbf{\bar H}_d$ are independent and identically distributed (i.i.d.) Gaussian random variables with zero mean and unit variance. The positive definite matrices $\mathbf \Psi_{\text{BS},\mathbf{G}} \in \mathbb C^{L \times L}$ and $\mathbf \Psi_{\text{BS},\mathbf{H}_d} \in \mathbb C^{L \times L}$ with unit diagonal entries denote the spatial correlation matrices at the BS seen from the reflected link $\mathbf{G}$ and the direct link $\mathbf{H}_d$, respectively. Similar definitions are employed for the correlation matrices at the UEs, i.e., $\mathbf \Psi_{\text{UE},\mathbf{H}_r} \in \mathbb C^{K \times K}$ and $\mathbf \Psi_{\text{UE},\mathbf{H}_d} \in \mathbb C^{K \times K}$, and the correlation matrices at the RIS, i.e., $\mathbf \Psi_{\text{RIS},\mathbf{H}_r} \in \mathbb C^{M \times M}$ and $\mathbf \Psi_{\text{RIS},\mathbf{G}} \in \mathbb C^{M \times M}$.
For simplicity, we set $\mathbf \Psi_{\text{BS},\mathbf{G}} = \mathbf \Psi_{\text{BS},\mathbf{H}_d} = \mathbf \Psi_\text{BS}$, $\mathbf \Psi_{\text{UE},\mathbf{H}_r} = \mathbf \Psi_{\text{UE},\mathbf{H}_d} = \mathbf \Psi_\text{UE}$, and $\mathbf \Psi_{\text{RIS},\mathbf{H}_r} = \mathbf \Psi_{\text{RIS},\mathbf{G}} = \mathbf \Psi_\text{RIS}$.
Then, we have
$
\text{vec} \left( \mathbf{H}_r  \right) \thicksim  \mathcal{CN}(\mathbf 0_{MK}, \mathbf \Psi_\text{UE} \otimes \mathbf \Psi_\text{RIS})$, $
\text{vec} \left(  \mathbf{G} \right) \thicksim  \mathcal{CN}(\mathbf 0_{LM}, \mathbf \Psi_\text{RIS} \otimes \mathbf \Psi_\text{BS})$, and $
\text{vec} \left(  \mathbf{H}_d \right) \thicksim  \mathcal{CN}(\mathbf 0_{LK}, \mathbf \Psi_\text{UE} \otimes \mathbf \Psi_\text{BS})
$ \cite{MatrixVariate}.
To proceed, based on the definition in (\ref{def:Gamma}), we have
\begin{align}
\mathbb E \left\{ \mathbf \Gamma_i^H \mathbf \Gamma_j\right\} &= \mathbb E \left\{ \mathbf{h}_{r,i} \mathbf{g}_{i}^H \mathbf{g}_j \mathbf{h}_{r,j}^H \right\} \nonumber\\
&= \mathbb E \left\{  \mathbf{g}_{i}^H \mathbf{g}_j \right\}  \mathbb E \left\{ \mathbf{h}_{r,i}\mathbf{h}_{r,j}^H \right\} \nonumber\\
&= L[ \mathbf \Psi_\text{RIS}]_{i,j} [\mathbf \Psi_\text{RIS} ]_{i,j} \mathbf \Psi_\text{UE} \nonumber \\
&= L [ \mathbf \Psi_\text{RIS} \odot \mathbf \Psi_\text{RIS} ]_{i,j} \mathbf \Psi_\text{UE},\ i \leq M, j \leq M.
\end{align}
On the other hand, it is easily seen that $\mathbb E \left\{ \mathbf \Gamma_{M+1}^H \mathbf \Gamma_{M+1}\right\} = \mathbb E \left\{ \mathbf{H}_d^H \mathbf{H}_d  \right\} = L \mathbf \Psi_\text{UE}$ and $\mathbb E \left\{ \mathbf \Gamma_i^H \mathbf \Gamma_j\right\} = \mathbf 0_{K \times K}$ when only one of $i$ and $j$ is equal to $M+1$ since $ \mathbf{\bar H}_d$ is independent of $\mathbf{\bar H}_r$ and $\mathbf{\bar G}$.
Therefore, the correlation matrix of the cascaded channel $\mathbf \Gamma$ is calculated as
\begin{align}
\mathbf R_{\mathbf \Gamma} =&\ \mathbb E \left\{ \begin{bmatrix} \mathbf \Gamma_1^H \\ \vdots \\ \mathbf \Gamma_{M+1}^H \end{bmatrix} [\mathbf \Gamma_1, \cdots, \mathbf \Gamma_{M+1} ]  \right\} \nonumber\\
=&\  \begin{bmatrix}
  L \left(\mathbf \Psi_\text{RIS} \odot \mathbf \Psi_\text{RIS} \right)  \otimes \mathbf \Psi_\text{UE}  & \mathbf 0_{KM \times K} \\
\mathbf 0_{K \times KM} &  L\mathbf \Psi_\text{UE}
\end{bmatrix}.
\end{align}

\section{Explanation on the MSE Performance Gap between LS and LMMSE Channel Estimators}\label{NMSEgap}
To explain the MSE performance gap in Figs.~\ref{fig:comparison} and \ref{fig:comparisonM} in mathematical terms, we focus on a scenario with a single UE and assume an uncorrelated channel model, i.e., $\mathbf R_{\mathbf \Gamma} = L\mathbf I_{M+1}$. In this case, considering an RIS with a unit amplitude response, it can be readily found that an orthogonal reflection pattern, i.e., $\mathbf V \mathbf V^H = (M+1)\mathbf I_{M+1}$, is optimal for both the LS and LMMSE channel estimators. The corresponding MSEs are given by
\begin{align}
J_\text{LS} =&\ L \text{Tr}\left[\left( \gamma (M+1)\mathbf I_{M+1} \right)^{-1}\right],
\nonumber\\
J_\text{LMMSE} =&\ L \text{Tr}\left[\left(\mathbf I_{M+1} + \gamma (M+1)\mathbf I_{M+1} \right)^{-1}\right],
\end{align}
respectively, where $ \gamma \triangleq \frac{P}{\sigma^2}$ denotes the SNR. As for the case with the phase reflection model in (\ref{fitness}), on the other hand, analytical solutions for $\mathbf V$ become elusive due to the intricate constraints imposed by the realistic model for the reflection coefficient. In this case, we denote the optimized $\mathbf V$ obtained through the proposed iterative algorithms using the LS and LMMSE channel estimation criteria, by $\mathbf V_\text{LS}^\star$ and $\mathbf V_\text{LMMSE}^\star$, respectively. Accordingly, the MSEs are given by
\begin{align}
J'_\text{LS} =&\ L \text{Tr}\left[\left( \gamma \mathbf V_\text{LS}^\star (\mathbf V_\text{LS}^\star)^H \right)^{-1}\right] ,
\nonumber\\
J'_\text{LMMSE} =&\ L \text{Tr}\left[\left(\mathbf I_{M+1} + \gamma \mathbf V_\text{LMMSE}^\star (\mathbf V_\text{LMMSE}^\star)^H \right)^{-1}\right].
\end{align}
In the sequel, we prove $\frac{J'_\text{LS}}{J'_\text{LMMSE}} > \frac{J_\text{LS}}{J_\text{LMMSE}}$ mathematically.

To begin with, we consider the following relationship:
\begin{align}\label{ieq1}
\frac{J'_\text{LS}}{J'_\text{LMMSE}}
&= \frac{\text{Tr}\left[\left( \gamma \mathbf V_\text{LS}^\star (\mathbf V_\text{LS}^\star)^H \right)^{-1}\right]}{\text{Tr}\left[\left(\mathbf I_{M+1} + \gamma \mathbf V_\text{LMMSE}^\star (\mathbf V_\text{LMMSE}^\star)^H \right)^{-1}\right]} \nonumber\\
&> \frac{\text{Tr}\left[\left( \gamma \mathbf V_\text{LS}^\star (\mathbf V_\text{LS}^\star)^H \right)^{-1}\right]}{\text{Tr}\left[\left(\mathbf I_{M+1} + \gamma \mathbf V_\text{LS}^\star (\mathbf V_\text{LS}^\star)^H \right)^{-1}\right]} \nonumber\\
&\triangleq  \frac{\text{Tr}\left[ \mathbf \Omega ^{-1}\right]}{\text{Tr}\left[\left(\mathbf I_{M+1} + \mathbf \Omega \right)^{-1}\right]},
\end{align}
where $\mathbf \Omega \triangleq \gamma \mathbf V_\text{LS}^\star (\mathbf V_\text{LS}^\star)^H$. The inequality is established since $\mathbf V_\text{LMMSE}^\star$ is the optimized solution for the LMMSE estimation and $\mathbf V_\text{LS}^\star$ lacks optimality under the LMMSE criterion. Next, we recall that better channel estimation performance is obtained if the RIS has a unit-amplitude response, i.e., $ J'_\text{LS} > J_\text{LS}$. Without loss of generality, we define $t \triangleq \frac{J'_\text{LS}}{J_\text{LS}} > 1$ and re-express $J'_\text{LS} = \text{Tr}\left[ \mathbf \Omega ^{-1}\right]$ as
\begin{align}\label{ieq2}
\text{Tr}\left[ \mathbf \Omega ^{-1}\right] = t J_\text{LS} &=  t \text{Tr}\left[\left( \gamma (M+1)\mathbf I_{M+1} \right)^{-1}\right] \nonumber\\
&= \text{Tr}\left[\left( \gamma/t  (M+1)\mathbf I_{M+1} \right)^{-1}\right].
\end{align}
To proceed, we need the following lemma.

\begin{lemma}\label{lemma2}
For an arbitrary $N \times N$ positive definite matrix $\mathbf A\succ \mathbf 0$ where $\text{Tr}[ \mathbf A^{-1}] = a = \text{Tr}[ (\frac{N}{a} \mathbf I_N)^{-1}]$, it holds that
\begin{align}
\text{Tr}[ (\mathbf I_N + \mathbf A)^{-1}] \leq  \text{Tr}\left[ \left(\mathbf I_N + \frac{N}{a} \mathbf I_N \right)^{-1} \right] = \frac{N a}{N + a}.
\end{align}
The equality holds if $\mathbf A = \frac{N}{a} \mathbf I_N$.
\end{lemma}
\begin{IEEEproof}
Denote the eigenvalues of $\mathbf A^{-1}$ by $\{t_n >0 \}_{n=1}^N$, which satisfy $\sum_{n=1}^N t_n = \text{Tr}[ \mathbf A ^{-1}] = a$. Then, we have
\begin{align}\label{f1}
\text{Tr}[ (\mathbf I_N + \mathbf A)^{-1}] = \sum_{n=1}^N \frac{1}{1+\frac{1}{t_n}} = N - \sum_{n=1}^N \frac{1}{1+t_n}.
\end{align}
By employing the Jensen's inequality to the convex function $f(x) = \frac{1}{1+x}$, we further have
\begin{align}\label{jensen}
\frac{1}{N}\sum_{n=1}^N \frac{1}{1+t_n} \geq  \frac{1}{1+ \frac{\sum_{n=1}^N t_n}{N}} =  \frac{N}{N+ a}.
\end{align}
Substituting (\ref{jensen}) into (\ref{f1}) yields the following relationship:
\begin{align}
\text{Tr}[ (\mathbf I_N + \mathbf A)^{-1}] \leq N - \frac{N^2}{N+ a} = \frac{N a}{N + a}.
\end{align}
The proof is completed.
\end{IEEEproof}

By applying the Lemma \ref{lemma2} to (\ref{ieq2}), we obtain
\begin{align}\label{ieq3}
 \text{Tr}\left[(\mathbf I_{M+1} + \mathbf \Omega) ^{-1}\right] &<  \text{Tr}\left[(\mathbf I_{M+1} + \gamma/t  (M+1)\mathbf I_{M+1}) ^{-1}\right] \nonumber\\
 &= t \text{Tr}\left[( t \mathbf I_{M+1} + \gamma  (M+1)\mathbf I_{M+1}) ^{-1}\right].
\end{align}
Accordingly, $\frac{\text{Tr}\left[ \mathbf \Omega ^{-1}\right]}{\text{Tr}\left[\left(\mathbf I_{M+1} + \mathbf \Omega \right)^{-1}\right]}$ fulfills the following properties:
\begin{align} \label{ieq4}
\frac{\text{Tr}\left[ \mathbf \Omega ^{-1}\right]}{\text{Tr}\left[\left(\mathbf I_{M+1} + \mathbf \Omega \right)^{-1}\right]} &=
\frac{t \text{Tr}\left[\left( \gamma (M+1)\mathbf I_{M+1} \right)^{-1}\right] }
{\text{Tr}\left[\left(\mathbf I_{M+1} + \mathbf \Omega \right)^{-1}\right]} \nonumber\\
&>
\frac{t \text{Tr}\left[\left( \gamma (M+1)\mathbf I_{M+1} \right)^{-1}\right] }
{t \text{Tr}\left[( t \mathbf I_{M+1} + \gamma  (M+1)\mathbf I_{M+1}) ^{-1}\right]} \nonumber\\
&> \frac{ \text{Tr}\left[\left( \gamma (M+1)\mathbf I_{M+1} \right)^{-1}\right] }
{ \text{Tr}\left[( \mathbf I_{M+1} + \gamma  (M+1)\mathbf I_{M+1}) ^{-1}\right]} \nonumber\\
&= \frac{J_\text{LS}}{J_\text{LMMSE}},
\end{align}
where the first equality is obtained by substituting (\ref{ieq2}), the second inequality follows from (\ref{ieq3}), and the
third inequality holds since $t > 1$.
Finally, by combining (\ref{ieq1}) and (\ref{ieq4}), we obtain $\frac{J'_\text{LS}}{J'_\text{LMMSE}} > \frac{J_\text{LS}}{J_\text{LMMSE}}$. Together with the fact $J'_\text{LMMSE} > J_\text{LMMSE}$, we conclude that, compared to the case study with an ideal unit amplitude reflection coefficient, the performance gap between the LS and LMMSE channel estimators becomes larger in the presence of a realistic model for the RIS reflection coefficient.

\end{appendices}

\end{document}